\documentclass[11pt,a4paper]{article}
\usepackage{jheppub,multirow}
\pdfoutput=1
\newcommand{\vev}[1]{{\left\langle {#1} \right\rangle}}

\DeclareMathOperator{\Tr}{Tr} 
\DeclareMathOperator{\tr}{tr} 
\DeclareMathOperator{\Spin}{Spin} 
\DeclareMathOperator{\SO}{SO} 
\DeclareMathOperator{\Sp}{Sp} 

\title{An Index for Ray Operators in 5d $E_n$ SCFTs}

\author{Chi-Ming~Chang,}
\author{Ori~J.~Ganor,}
\author{and Jihwan Oh}

\affiliation{
Department of Physics,
  University of California,\\
Berkeley, CA 94720, U.S.A.}

\emailAdd{cmchang@berkeley.edu}
\emailAdd{ganor@berkeley.edu}
\emailAdd{jihwanoh@berkeley.edu}

\abstract{
We construct an index for BPS operators supported on a ray in five dimensional superconformal field theories with exceptional global symmetries. 
We compute the $E_n$ representations (for $n=2,\dots,7$) of operators of low spin, thus verifying that while the expression for the index is only SO$(2n-2)\times$U(1) invariant, the index itself exhibits the full $E_n$ symmetry (at least up to the order we expanded).
The ray operators we studied in 5d can be viewed as generalizations of operators constructed in a Yang-Mills theory with fundamental matter by attaching an open Wilson line to a quark. For $n\le 7$, in contrast to local operators, they carry nontrivial charge under the $\mathbb{Z}_{9-n}\subset E_n$ center of the global symmetry.
The representations that appear in the ray operator index are therefore different, for $n\le 7$, from those appearing in the previously computed superconformal index. 
For $3\le n\le 7$, we find that the leading term in the index is a character of a minuscule representation of $E_n$.
We also discuss the case $n=8$, which presents a unique technical challenge, and remains an open problem.
}

\begin{document}
\maketitle
\flushbottom

\newcommand{\secref}[1]{\S\ref{#1}}
\newcommand{\figref}[1]{Figure~\ref{#1}}
\newcommand{\appref}[1]{Appendix~\ref{#1}}
\newcommand{\apprefrange}[2]{Appendices~\ref{#1}-\ref{#2}}
\newcommand{\tabref}[1]{Table~\ref{#1}}

\newcommand\rep[1]{{\bf {#1}}} 
\newcommand\brep[1]{{\overline{\bf {#1}}}} 

\newcommand\SUSY[1]{{${\mathcal{N}}={#1}$}}  
\newcommand\px[1]{{\partial_{#1}}}

\def\be{\begin{equation}}
\def\ee{\end{equation}}
\def\bear{\begin{eqnarray}}
\def\eear{\end{eqnarray}}
\def\nn{\nonumber}
\def\ie{\begin{equation}\begin{aligned}}
\def\fe{\end{aligned}\end{equation}}

\newcommand\bra[1]{{\left\langle{#1}\right\rvert}} 
\newcommand\ket[1]{{\left\lvert{#1}\right\rangle}} 

\newcommand{\C}{\mathbb{C}}
\newcommand{\R}{\mathbb{R}}
\newcommand{\Z}{\mathbb{Z}}
\newcommand{\CP}{\mathbb{CP}}

\def\Id{{\mathbf{I}}} 

\def\SO{{\rm SO}}
\def\SU{{\rm SU}}

\def\Mst{M_{\text{st}}} 
\def\gst{g_{\text{st}}} 
\def\gIIB{g_{\text{IIB}}} 
\def\lst{\ell_{\text{st}}} 
\def\lP{\ell_{\text{P}}} 

\def\xR{{R}}

\def\gYM{g_{\text{ym}}}

\def\ten{{\natural}} 

\def\elDirac{{\Gamma}} 
\def\tenDirac{{\gamma}} 

\def\uE{{\mu}} 
\def\vE{{\nu}} 

\def\qk{{\mathbf{u}}}
\def\bqk{{\overline{\qk}}}

\def\iF{{\mathbf{i}}} 
\def\jF{{\mathbf{j}}} 

\def\PathP{{\mathcal{P}}} 

\def\SurfS{{\mathcal{S}}} 
\def\LoopC{{\mathcal{C}}} 

\def\nC{{n}}
\def\aC{{\alpha}}

\def\pSUSY{{\varepsilon}}

\def\hE{{\widehat{E}}}

\def\xV{{\mathbf{a}}}

\def\tw{{\gamma}}

\def\gS{{\mathbf{h}}} 
\def\rS{{\rho}} 
\def\xS{{\mathbf{y}}} 
\def\iS{{\alpha}} 
\def\jS{{\beta}} 
\def\kS{{\gamma}} 
\def\lS{{\delta}} 

\def\tS{{\tau}} 

\def\SP{{SP}}
\def\SPole{{\mathbf{p}}}
\def\NP{{NP}}
\def\NPole{{\mathbf{q}}}

\def\gLam{{\Lambda}}
\def\gGaugeAll{{\mathbf{G}}} 

\def\gGauge{{\widetilde{\mathbf{G}}}} 

\def\kInst{{\mathbf{k}}} 

\def\ModInst{{\widetilde{{\mathcal M}}}} 

\def\Zt{{\tilde{Z}}}
\def\ft{{\tilde{f}}}

\def\Ray{{\mathfrak{R}}} 
\def\WR{{\mathcal{R}}} 

\def\xTB{$\times$} 

\def\OCCA{{\mathfrak{V}}} 

\def\aCounter{{\mathfrak{a}}}
\def\aSet{{\mathbf{A}}}

\def\aALA{{\mathbf{a}}}

\def\Adt{{\mathfrak{A}}} 

\def\SpA{{A}} 
\def\SpPhi{{\Phi}} 

\def\qk{{q}} 
\def\bqk{{\overline{\qk}}} 
\def\iF{{\mathbf{i}}} 
\def\covD{{D}} 
\def\cLO{{{\mathcal O}}} 

\def\qQ{{\mathbf{Q}}} 

\def\UoneJ{{\mathbf{j}}} 
\def\UoneA{{\mathbf{a}}} 
\def\UoneF{{\mathbf{f}}} 
\def\UonePhi{{\varphi}} 

\def\PhiVEV{{v}} 

\def\mFlux{{\mathbf{m}}} 
\def\mIIAM{{\mathbf{M}}} 
\newcommand\aDp[1]{{{\mathbf{a}}_{(\text{D}{#1})}}} 
\newcommand\fDp[1]{{{\mathbf{f}}_{(\text{D}{#1})}}} 
\newcommand\latQrt[1]{{Q_{\text{rt}}^{({#1})}}} 
\newcommand\latQwt[1]{{Q_{\text{wt}}^{({#1})}}} 

\def\kCS{{\tilde{k}}} 
\def\nStrings{{{n_1}}} 

\def\defineas{{\stackrel{\mbox{\tiny{def}}}{=}}}

\def\SinhArg{{\mathbf{X}}} 
\newcommand\Zvec[1]{{{\mathcal Z}_{\text{\tiny vec}}^{{#1}}}}
\newcommand\Zanti[1]{{{\mathcal Z}_{\text{\tiny anti}}^{{#1}}}}
\newcommand\Zfund[1]{{{\mathcal Z}_{\text{\tiny fund}}^{{#1}}}}
\newcommand\Zprime[1]{{{\mathcal Z}_{\text{\tiny prime}}^{{#1}}}}

\def\eP{{\epsilon_{+}}}
\def\eM{{\epsilon_{-}}}

\def\JKQ{{\mathbf{Q}}} 
\def\JKeta{{\eta}} 

\def\latQ{{\mathbf{Q}}} 

\newcommand\innerP[2]{{({{#1}}|{{#2}})}} 

\def\Zcenter{{\mathcal{Z}}} 

\def\lStrings{{l}} 

\def\ZinstLine{{{\cal Z}_{\text{\tiny inst}+\text{\tiny line}}}}

\section{Introduction}
\label{sec:Intro}

There is strong evidence for an interacting 5d superconformal field theory (SCFT) with $E_8$ global symmetry and a one-dimensional Coulomb branch \cite{Seiberg:1996bd}. A few of its (dual) realizations in string theory are the low energy limits of the systems listed below:
\begin{enumerate}
\item[(i)]
A D$4$-brane probing a 9d $E_8$ singularity in type-I' string theory \cite{Seiberg:1996bd}; the latter is realized by the infinitely strong coupling limit of seven coincident D$8$-branes and an orientifold (O$8$) plane \cite{Polchinski:1995df}.

\item[(ii)]
M-theory on a certain degenerate Calabi-Yau manifold \cite{Douglas:1996xp,Intriligator:1997pq,Ganor:1996pc}; the Calabi-Yau threefold can be taken as the canonical line bundle of a del Pezzo surface $B_8$ (which can be constructed as the blow-up of $\CP^2$ at $8$ points) in the limit that the volume of $B_8$ goes to zero. (See also \cite{Witten:1996qb} where the study of such a limit of M-theory was initiated, and \cite{Morrison:1996na,Morrison:1996pp} where the F-theory version of this degeneration was described.)

\item[(iii)]
The 6d $E_8$ SCFT \cite{Ganor:1996mu,Seiberg:1996vs} compactified on S$^1$ \cite{Ganor:1996pc}.

\item[(iv)]
Webs of $(p,q)$ $5$-branes \cite{Aharony:1997bh}.

\end{enumerate}

The $E_8$ theory can be deformed by relevant operators to 5d SCFTs with smaller $E_n$ global symmetries ($n=0,\dots,7$).
One of the remarkable achievements of the last few years has been the construction of a supersymmetric index that counts local operators that preserve (at least) $\frac{1}{8}$ of the supersymmetry of the $E_n$ theories \cite{Kim:2012gu,Iqbal:2012xm,Bergman:2013ala,Hayashi:2013qwa,Bao:2013pwa,Hwang:2014uwa}.

Technically, this {\it superconformal index} is constructed by computing the partition function on S$^4$$\times$S$^1$ of a 5d supersymmetric gauge theory with gauge group SU(2) and $N_f=n-1$ hypermultiplets. The global flavor symmetry is SO($2N_f$), which combines with the U(1) instanton charge to form the SO($2n-2)\times$U(1) $\subset E_n$, as predicted in \cite{Seiberg:1996bd}. The partition function is computed using the techniques developed in \cite{Pestun:2007rz}, with insight from string theory for the proper treatment of zero size instantons \cite{Witten:1995gx}. The partition function is presented as an integral over a product of Nekrasov partition functions \cite{Nekrasov:2002qd}, and the resulting index is expressed as an infinite sum of monomials in fugacities that capture spin and R-charge, with coefficients that are linear combinations of characters of SO($2n-2)\times$U(1). It is remarkable that these linear combinations of characters match representations of $E_n\supset$ SO($2n-2)\times$U(1).

The string-theory or M-theory realizations of the $E_n$ theories allow for a construction of BPS line operators akin to Wilson lines as follows. In the type-I' setting (i), we introduce a semi-infinite fundamental string (F$1$) perpendicular to the plane of the D$8$-branes with one of its endpoints at infinity and the other on the D$4$-brane.
In the M-theory setting (ii), we add an M$2$-brane that fills the $\C$ fiber of the canonical bundle above a point of the del Pezzo base. In the 6d setting (iii), the line operator is the low-energy limit of a surface operator on $S^1$, and in the type-IIB setting (iv), it is realized by an open $(p,q)$ string.
In addition, the 5d $E_n$ theories also possess BPS operators supported on a line with an endpoint, which we will refer to as {\it Ray operators.}
They are analogous to a Wilson line along a ray, capped by a quark field at the endpoint.
The aim of this paper is to study these 5d ray operators and extend the results of \cite{Kim:2012gu,Hwang:2014uwa} by constructing an index for $\frac{1}{8}$BPS ray operators.

Calculating the index for ray operators again requires a careful treatment of zero-size instantons and additional insight from string theory. The index can again be written in terms of characters of SO($2n-2)\times$U(1) which combine into characters of $E_n$. 
Unlike local operators, the ray operators are charged under the center of $E_n$, in the cases where it is nontrivial ($n<8$). The appearance of complete $E_n$ characters is a nontrivial check of the validity of the assumptions behind the computation of the index. Moreover, for $n<8$ the weight lattice is larger than the root lattice of $E_n$, and we find $E_n$ representations that do not appear in the superconformal index. For example, for $E_6$ we find the representations $\rep{27}$, $\rep{1728}$, etc., consistent with the $\Z_3$ charge of the ray operator.

Our paper is organized as follows.
In \secref{sec:Rev5dEnIndex} we review the construction of 5d SCFTs with $E_n$ global symmetry and their superconformal indices. In \secref{sec:LRO} we introduce ray operators into the 5d SCFTs, and we compute their indices in \secref{sec:IndexRO} (with our final results in \secref{subsec:IndexRays}). 
We conclude with a summary and discussion in \secref{sec:Disc}.

\section{Review of the 5d $E_n$ SCFTs and their superconformal indices}
\label{sec:Rev5dEnIndex}

Following the discovery of \cite{Polchinski:1995df} that in the infinite string coupling limit of type I' string theory, a 9d $E_n$ gauge theory describes the low-energy limit of $N_f=n-1$ D$8$-branes coincident with an O$8$-plane, Seiberg constructed a 5d SCFT with $E_n$ global symmetry by probing the D$8$/O$8$ singularity with $N$ D$4$-branes \cite{Seiberg:1996bd}. The brane directions are listed in the table below.
\begin{center}
\begin{tabular}{|c|cccccccccc|} \hline
   &  0 &  1 &  2 &  3 &  4 &  5 &  6 &  7 &  8 &  9 \\
\hline\hline
D8/O8 &\xTB&\xTB&\xTB&\xTB&\xTB&\xTB&\xTB&\xTB&\xTB&    \\
\hline
D4 &\xTB&\xTB&\xTB&\xTB&\xTB&    &    &    &    &    \\
\hline
\end{tabular}
\end{center}
Denoting by $\epsilon_L$, $\epsilon_R$  left and right 10d Majorana-Weyl spinors, the configuration preserves those SUSY parameters that satisfy $\Gamma^{012345678}\epsilon_L=\Gamma^{01234}\epsilon_L=\epsilon_R$. The rotations in directions $5\dots 8$ act on spinors as SU(2)$_+^R\times$SU(2)$_-^R$. The second factor acts trivially on the supercharges, while the first factor acts nontrivially and is identified with the R-symmetry SU(2)$_R$ of the 5d theory. The theory has an $N$-dimensional Coulomb branch $(\R_{+})^N/S_N$ that can be identified with the D$4$-branes moving away from the $E_n$ singularity, and a Higgs branch that can be identified with the moduli space of $E_n$ instantons at instanton number $N$. The supermultiplet associated with the center of mass of the D$4$-branes in directions $5\dots 8$ decouples and is not considered part of the SCFT.

Raising the value of the inverse string coupling constant $1/\gst$ at the common position of the D$8$-branes and O$8$-plane from zero (formally $\gst=\infty$) to a nonzero value breaks the global $E_n$ symmetry to SO($2N_f)\times$U(1). The SO($2N_f$) factor comes from the gauge symmetry of the D$8$-branes, while the U(1) factor is associated with D$0$-brane charge.
At low-energy the D$4$-brane probe theory is then described by $\Sp(N)$ SYM coupled to $N_f$ hypermultiplets in the fundamental representation $\rep{2N}$ of the gauge group, and also a single hypermultiplet in the antisymmetric representation. SO($2N_f)$ is the global flavor symmetry, while U(1) is the symmetry associated with instanton number whose conserved current is
\ie\label{InstantonCurrent}
J={1\over 8\pi^2}\tr\star(F\wedge F).
\fe
Here $F$ is the $\Sp(N)$ field strength.

In this paper, we will focus on the Sp(1) gauge theories, which have no antisymmetric hypermultiplet. The vector multiplet consists of a gauge field $A_\mu$, a real scalar $\SpPhi$, and symplectic-Majorana fermions $\lambda^A_m$, where $A=1,2$ denotes the SU(2)$_R$ R-symmetry doublet index, and $m=1,\ldots,4$ denotes the SO(1,4) spinor index. The hypermultiplet consists of complex scalars $q^A$ and fermions $\psi_m$.

\subsection{Superconformal indices of 5d SCFTs}
\label{subsec:Index}

We will now discuss some general aspects of the superconformal index  \cite{Bhattacharya:2008zy} of 5d SCFTs. The superconformal algebra of 5d SCFTs is $F(4)$, which in Euclidean signature has the bosonic subgroup SO(1,6)$\times$SU(2)$_R$.  One of the most important consequences of the superconformal symmetry is the BPS bound, which follows from the anticommutator \cite{Bhattacharya:2008zy},
\ie\label{SCA}
\{Q^A_m,S^n_B\}=\delta^n_m \delta^A_B D+2\delta^A_B M_m{}^n-3\delta^n_mR_B{}^A,
\fe
where $D$ is the dilatation generator, $M_m{}^n$ are the SO(5) rotation generators, and $R_B{}^A$ are the SU(2)$_R$ R-symmetry generators. $Q^A_m$ and $S^m_A$ are the supercharge and superconformal charge. Note that $M_m{}^n$ are components of an $\Sp(2)$ matrix. We will work in the basis where $M_1{}^1=-M_2{}^2$ and $M_3{}^3=-M_4{}^4$, and we will denote $J_+\equiv M_1{}^1$, $J_-\equiv M_3{}^3$, which are the Cartan generators of SU(2)$_+\times$SU(2)$_-\subset$ SO(5). We also denote $J_R\equiv R_1{}^1$, which is the Cartan generator of SU(2)$_R$. 
In radial quantization, the superconformal generator is the hermitian conjugate of the supercharge, i.e., $S^m_A=(Q^A_m)^\dagger$. The anticommutator \eqref{SCA} implies positivity conditions on linear combinations of the dilatation, rotations, and R-symmetry generators. 
For instance, in the case of $m=2$ and $A=1$, \eqref{SCA} implies
\be\label{BPSbound}
\Delta\equiv \{Q,S\}=D - 2J_+ - 3 J_R\ge 0,
\ee
where for simplicity we denote $Q\equiv Q_2^1$, $S\equiv S^2_1$. The operators that saturate the BPS bound \eqref{BPSbound} are called $\frac{1}{8}$BPS operators. These operators are annihilated by both $Q$ and $S$. By the state-operator correspondence, the space of local operators is isomorphic to the Hilbert space ${\cal H}$ of the (radially quantized) theory on S$^4$. The number of $\frac{1}{8}$BPS operators with given quantum numbers (counted with $\pm$ signs according to whether they are bosonic or fermionic) is captured by the {\it superconformal index},
\be\label{SuperconformalIndex}
{\cal I}_{\text{\tiny SCI}}(\epsilon_+,\epsilon_-,m_i)=\Tr_{\cal H}\left[(-1)^F e^{-\beta\Delta}e^{-2\epsilon_+ (J_{+} +J_R)-2\epsilon_- J_{-} } e^{-\sum F_i m_i}\right],
\ee
where $F$ is the fermion number operator, and $F_i$ denote the generators of other global symmetries.\footnote{In this paper, we will sometimes parametrize the indices using the fugacities $t=e^{-\epsilon_+}$ and $u=e^{-\epsilon_-}$.} Only the states that saturate the BPS bound (with $\Delta=0$) contribute to the trace, and the contributions from states with nonzero $\Delta$ pairwise cancel out due to $(-1)^F$, since $Q$ and $S$ commute with the other operators inside the trace.

The $\frac{1}{8}$BPS operators are annihilated by both the  supercharge $Q$ and one superconformal charge $S$. Formally, if we regard $Q$ as an exterior derivative $d$ and $S$ as its Hermitian conjugate $d^\star$, then $\{Q,S\}$ corresponds to the Laplacian $\Delta=d^\star d+d d^\star$. Hodge theorem states that the space of harmonic forms (states with $\Delta = 0$) is isomorphic to the cohomology of $d$.  Analogous arguments, formulated in terms of $Q$, $S$, show that the Hilbert space ${\cal H}$ of $\frac{1}{8}$BPS operators is isomorphic to the cohomology of $Q$ \cite{Grant:2008sk}, which will be referred to as $Q$-cohomology. The superconformal index can be interpreted as the Euler characteristic of the $Q$-cohomology.\footnote{The $Q$-cohomology contains more information than its Euler characteristic. For example, the Hilbert-Poincar\'e polynomial of the $Q$-cohomology gives the partition function of the BPS operators
\cite{Grant:2008sk,Chang:2013fba}.}

Consider an SU(2)$_{+}\times$SU(2)$_R$ multiplet of operators with SU(2)$_{+}$ spin $j_{+}$ and SU(2)$_R$ spin $j_R$. Out of the $(2j_{+}+1)(2j_R+1)$ states, at most one can saturate the BPS bound \eqref{BPSbound} --- this is the state with maximal $J_{+}=j_{+}$ and $J_R=j_R$.
Thus, a $\tfrac{1}{8}$BPS state that contributes to the index \eqref{SuperconformalIndex} has maximal $J_{+}$ and $J_R$ charge in its SU(2)$_{+}\times$SU(2)$_R$ multiplet.
Consider a $\tfrac{1}{8}$BPS state with $J_R=0$. According to the above discussion, it must be a singlet of SU(2)$_R$. The algebra of SU(2)$_R$ is generated by $R_1{}^1$, $R_1{}^2$ and $R_2{}^1$, and since the state is annihilated by both $R_1{}^2$ and $Q\equiv Q_2^1$, it must be annihilated by $[R_1{}^2,Q_2^1]$ which is proportional to $Q_2^2$. Similarly, we see that it is annihilated by $S_1^1$ as well. It therefore has enhanced supersymmetry, being annihilated by $Q_2^1$, $Q_2^2$, $S_1^2$, $S_2^2$, and is in fact a $\tfrac{1}{4}$BPS.
Similarly,  a $\tfrac{1}{8}$BPS state with $J_{+}=0$ is a singlet of SU(2)${}_{+}$ and is annihilated by $M_1{}^1$, $M_1{}^2$ and $M_2{}^1$, and therefore also by $[M_1{}^2,Q_2^1]$ and $[M_2{}^1,S_1^2]$. It is thus annihilated by $Q_1^1$, $Q_2^1$, $S_1^1$, $S_1^2$. Moreover, since it saturates the BPS bound $D=3J_R$, it must be an SU(2)${}_{-}$ singlet as well. This is because similarly to the BPS bound \eqref{BPSbound}, we also have in general
\be\label{eqn:BPSboundQ4S4}
\{Q_4^1,S_1^4\}=D - 2J_{-} - 3 J_R\ge 0.
\ee
If the state in question had nonzero SU(2)${}_{-}$ spin $j_{-}$, then it would be part of a multiplet of $(2j_{-}+1)$ states with $D=3J_R$, but the state with maximal $J_{-}=j_{-}$ in that multiplet would then violate the bound \eqref{eqn:BPSboundQ4S4}.
It follows that a  $\tfrac{1}{8}$BPS state with $J_{+}=0$ must also have $J_{-}=0$ and is in fact $\tfrac{1}{2}$BPS, being annihilated by all $Q_m^1$ and all $S_1^m$ ($m=1,\dots,4$).
For example, a $\tfrac{1}{8}$BPS state with $J_{+}+J_R=\tfrac{1}{2}$ must have either $J_{+}=0$ or $J_R=0$ and is therefore at least $\tfrac{1}{4}$BPS. If it is not an SU(2)${}_{-}$ singlet, then it must have $J_R=0$ and $J_{+}=\tfrac{1}{2}$ and is  $\tfrac{1}{4}$BPS. As another example,
a $\tfrac{1}{8}$BPS state that has $J_{+}+J_R=0$ is a singlet of both SU(2)$_{+}$ and SU(2)$_R$. It is therefore annihilated by all $Q_m^A$ and $S_A^m$ and must be the vacuum state.
 Thus in an expansion of the index \eqref{SuperconformalIndex} in $e^{-2\eP}$, the only term that is $\eP$-independent is the contribution of the identity operator $1$.
Other terms can be expanded in characters of SU(2)${}_{-}$,
$$
\chi_{2j+1}(e^{-\eM})=
\sum_{m=-j}^{j} e^{-2m\eM}
=\frac{\sinh(2j+1)\eM}{\sinh\eM}\,.
$$
The terms linear in $e^{-\eP}$ and proportional to $\chi_{2j+1}(e^{-\eM})$, with $j>0$, are the contributions of  $\tfrac{1}{4}$BPS states.
 Note that even when both $J_{+}$ and $J_R$ are nonzero, the $1\over 8$BPS states preserve half of the supercharges, namely $Q^1_2$ as well as $Q^1_1$, $Q^1_3$, and $Q^1_4$, since the latter three lower $\Delta$ (but the hermitian conjugates of $Q^1_1$, $Q^1_3$, and $Q^1_4$ are in general not preserved).

In Euclidean signature, the space ${\mathbb R}^5$ can be conformally mapped to ${\mathbb R}\times$S$^4$, and the superconformal index \eqref{SuperconformalIndex} can be interpreted as a twisted partition function of the theory on S$^1\times$S$^4$. For theories with a Lagrangian description, it can be computed by a path integral with the fields satisfying periodic boundary conditions along  S$^1$, further twisted by the various fugacities.

\subsection{Superconformal indices from 5d SYM}
\label{subsec:SCIfromSYM}

Now, let us focus on the $E_n$ SCFTs. It has been shown that the superconformal indices of them can be computed using the IR 5d SYM with fundamental matters \cite{Kim:2012gu,Hwang:2014uwa}.

In the IR theory, the superconformal algebra is not defined, because the Yang-Mills coupling constant is dimensionful, but we can consider the $Q$-cohomology on gauge invariant operators.

A large class of gauge invariant operators can be constructed from the `letters' of the 5d ${\cal N}=1$ gauge theory, given by the fields listed in the following table, acted on by an arbitrary numbers of derivatives, modulo the free field equations of motion.
\begin{center}
\begin{tabular}{|c|c|c|c|c|c|c|c|c|}\hline
 & $F_{\mu\nu}$ & $\lambda^A_m$ & $\SpPhi$ & $q^A$ & $\psi_m$ & $\partial_\mu$ & $Q$ 
\\ \hline 
$\widetilde E$ & 2 & ${3\over 2}$ & $1$ & ${3\over 2}$ & $2$ & $1$ & ${1\over 2}$ 
\\ \hline
\end{tabular}
\end{center}

In this table, we also introduced a new quantum number $\widetilde E$ for the fields, the derivative symbol and the supercharge $Q$, and we define
\ie
\widetilde\Delta=\widetilde E-2J_+ - 3J_R,
\fe
so that $\widetilde\Delta(Q)=0$\footnote{Recall $Q\equiv Q^1_2$ has $J_+=-1/2$ and $J_R=1/2$.}, which will help with computing the $Q$-cohomology. One should not confuse $\widetilde E$ and $\widetilde\Delta$ with the dimension $D$ and the radial Hamiltonian $\Delta$ that appear in the BPS bound formula \eqref{BPSbound}. We emphasize that $\widetilde E$, which measures the classical dimension of the corresponding field in the SYM theory, is purely a bookkeeping device and the $Q$-cohomology does not depend on the assignment of the $\widetilde E$. Let us first consider the single-letter $Q$-cohomology. The supersymmetry transformation on the component fields in the vector and hypermultiplets can be found in (2.10) and (2.14) of \cite{Kim:2012gu}. It is not hard to see that the operators with $\widetilde \Delta\ge1$ have trivial $Q$-cohomology. On the other hand, for $\widetilde \Delta = -1$ and $0$, we have nontrivial cohomology generated by
\ie
\widetilde \Delta=-1:~~~&\lambda_{+0+},
\\
\widetilde \Delta=0:~~~&\lambda_{0\pm+},~~q^+,
\fe
and also the two derivatives $\partial_{+\pm}$ acting on them. The subscripts of $\partial_{\pm\pm}$ and the first two subscripts of $\lambda_{\pm0\pm}$, $\lambda_{0\pm\pm}$ denote their $2J_+$ and $2J_-$ charges. The last subscripts of $\lambda_{\pm0\pm}$, $\lambda_{0\pm\pm}$ and also the superscript of $q^+$ denote their $2J_R$ charges.\footnote{$q^\pm=q^1$.}
 Note that only those components with maximal $J_{+}$ and $J_R$ in an SU(2)${}_{+}\times$SU(2)${}_R$ multiplet, for each field, can be generators of a nontrivial $Q$-cohomology.

The single-letter operators are subject to an equation of motion,
\ie
\partial_{++}\lambda_{0-+} +\partial_{+-}\lambda_{0++}
=-\partial_5\lambda_{+0+},
\fe
and it is not hard to check that $\partial_5\lambda_{+0+}$ is $Q$-exact, and therefore vanishes in the $Q$-cohomology.
We compute the single-letter index by summing over the letters ($F$, $\lambda$, $\Phi$, $q$, $\psi$):
\ie
f=\sum_{\rm letters}(-1)^F t^{2(J_++J_R)}u^{2J_-}e^{-\sum F_i m_i}=f_{\text{\tiny\bf adj}}+f_{\text{\tiny\bf fund}},
\fe
where $t$ and $u$ are related to $\epsilon_+$ and $\epsilon_-$ by $t=e^{-\epsilon_+}$ and $u=e^{-\epsilon_-}$, the $m_i$ ($i=1,\dots,N_f$) are the chemical potentials of the flavor charges of a Cartan subalgebra U(1)${}^{N_f}\subset$ O($2N_f$),
and the single-letter index for the vector multiplet and fundamental hypermultiplet are given by
\ie\label{SingleLetterPF}
&f_{\text{\tiny\bf adj}}={-t^2-t(u+u^{-1})+t^2\over (1-tu)(1-tu^{-1})}=-{t(u+u^{-1})\over (1-tu)(1-tu^{-1})},
\\
&f_{\text{\tiny\bf fund}}={t\over (1-tu)(1-tu^{-1})}\sum_{\ell=1}^{N_f}2\cosh m_\ell.
\fe
The multi-letter index can be computed by the following formula \cite{Aharony:2003sx,Kinney:2005ej},
\ie\label{PertIndex}
&{\cal I}(t,u,m_i)=\int_{\rm Sp(1)} {\cal Z}_{\text{\tiny1-loop}}(t,u,w,m_i)dU,
\\
&{\cal Z}_{\text{\tiny1-loop}}(t,u,w,m_i)=\exp\bigl[
\sum_{\substack{{\bf R}\\ \text{\{\tiny \bf fund, \bf adj}\}\\}}
\sum_{n=1}^\infty {1\over n}f_{\bf R}(t^n,u^n,m_i^n)\chi_{\bf R}(w^n)\bigr],
\fe
where the integral is over the Sp(1) matrices $U$, and $w$ is one of the eigenvalues of $U$.  The $U$-integral can be simplified to a one-dimensional integral $dU={1\over \pi}\sin^2\alpha d\alpha$ where $w=e^{i\alpha}$. The representations $\bf R$ that appear in the sum are the fundamental and adjoint representations.  $\chi_{\bf R}(w)$ are the Sp(1) characters, for example $\chi_{\text{\tiny\bf adj}}(w)=w^{-2}+1+w^2$ and $\chi_{\text{\tiny\bf fund}}(w)=w^{-1}+w$. One can recognize that the integrand ${\cal Z}_{\text{\tiny1-loop}}$ is a multi-letter index that counts gauge covariant operators, and the integration over the gauge group imposes the gauge invariance. The single-letter indices \eqref{SingleLetterPF} and the formula \eqref{PertIndex} can also be derived by evaluating a path integral of the 5d SYM on S$^1\times$S$^4$ \cite{Pestun:2007rz,Kim:2012gu}, where the matrix $U$ is identified with the Sp(1) holonomy along the S$^1$.

The index \eqref{PertIndex} cannot be the full superconformal index, because all the gauge invariant operators that contribute to \eqref{PertIndex} do not carry the topological U(1) charge associated with the conserved current \eqref{InstantonCurrent}.

The contributions of the operators with $n$ units of the topological U(1) charge to the superconformal index can be computed in the path integral on S$^1\times$S$^4$ with the field strength restricted to the $n$-th instanton sector,
\ie
{1\over 8\pi^2}\int_{{\rm S}^4}\tr (F\wedge F) = n.
\fe
In \cite{Pestun:2007rz,Kim:2012gu}, using supersymmetric localization, it was shown that the path integral localizes at the singular instanton solution at the south pole and anti-instanton solution at the north pole. Near the south (north) poles, the spacetime looks like S$^1\times{\mathbb R}^4$, and the path integral over the solutions to the instanton (anti-instanton) equation reduces to the the Nekrasov instanton partition function ${\cal Z}_{\text{\tiny inst}}(t,u,m_i,q)$ in the $\Omega$-background on ${\mathbb R}^4$. The superconformal index is then computed by the formula
\be\label{SCI}
{\cal I}_{\text{\tiny SCI}}(t,u,m_i,q)=\int _{\rm Sp(1)} {\cal Z}_{\text{\tiny1-loop}}(t,u,w,m_i) |{\cal Z}_{\text{\tiny inst}}(t,u,w,m_i,q)|^2dU.
\ee
where ${\cal Z}_{\text{\tiny inst}}(t,u,w,m_i,q)^*={\cal Z}_{\text{\tiny inst}}(t,u,w^{-1},-m_i,q^{-1})$ is the contribution from the anti-instantons at the north pole.

In \cite{Hwang:2014uwa}, it has been argued that the Nekrasov instanton partition function can be computed by the Witten indices of certain D0-brane quantum mechanics. In the next subsection, we review the D0-brane quantum mechanics, and compute their Witten indices.

\subsection{The D0-D4-D8/O8 system}
\label{sec:D0-D4-D8/O8}

\begin{table}\centering
\begin{tabular}{|c|c|c|c|}
\hline
strings &${\cal N}=4$ multiplets & fields & {\footnotesize SU(2)$_-\times$SU(2)$_+\times$SU(2)$^R_-\times$SU(2)$^R_+$} 
\\ \hline
\multirow{8}{*}{D0-D0 strings}& \multirow{3}{*}{vector} & gauge field &$({\bf 1},{\bf 1},{\bf 1},{\bf 1})$ 
\\ \cline{3-4}
& & scalar & $({\bf 1},{\bf 1},{\bf 1},{\bf 1})$ 
\\ \cline{3-4}
&  & fermions &$({\bf 1},{\bf 2},{\bf 1},{\bf 2})$ 
\\ \cline{2-4}
&Fermi & fermions & $({\bf 2},{\bf 1},{\bf 2},{\bf 1})$
\\ \cline{2-4}
&\multirow{2}{*}{twisted hyper}  & scalars & $({\bf 1},{\bf 1},{\bf 2},{\bf 2})$ 
\\ \cline{3-4}
&& fermions & $({\bf 1},{\bf 2},{\bf 2},{\bf 1})$ 
\\ \cline{2-4}
&\multirow{2}{*}{hyper} & scalars & $({\bf 2},{\bf 2},{\bf 1},{\bf 1})$ 
\\ \cline{3-4}
&& fermions & $({\bf 2},{\bf 1},{\bf 1},{\bf 2})$ 
\\ \hline
\multirow{3}{*}{D0-D4 strings}& \multirow{2}{*}{hyper} & scalars & $({\bf 1},{\bf 2},{\bf 1},{\bf 1})$ 
\\ \cline{3-4}
& & fermions & $({\bf 1},{\bf 1},{\bf 1},{\bf 2})$ 
\\ \cline{2-4}
& Fermi & fermions & $({\bf 1},{\bf 1},{\bf 2},{\bf 1})$ 
\\ \hline
D0-D8 strings & Fermi & fermions & $({\bf 1},{\bf 1},{\bf 1},{\bf 1})$ 
\\ \hline
\end{tabular}
\caption{The field content of the D0-D4-D8/O8 quantum mechanics.}
\label{table:D0-D4-D8/O8}
\end{table}

The instantons in the IR 5d Sp(1) SYM of the $E_n$ theory are described by the D0-branes moving in the background of one D4-brane and $N_f$ D8-branes coincident with an O8-plane \cite{Hwang:2014uwa}. 
The low energy theory on $k$ D0-branes is a ${\cal N}=4$ O($k$) gauged quantum mechanics, whose field content is listed in 
\tabref{table:D0-D4-D8/O8}, where the last column lists the representations of various fields under the R-symmetry SU(2)$_+\times$SU(2)$^R_+$ and global symmetry SU(2)$_-\times$SU(2)$^R_-$. The SU(2)$_+\times$SU(2)$_-$ and SU(2)$^R_+\times$SU(2)$^R_-$ are the rotation groups of the four-planes $\R^{1234}$ and $\R^{5678}$, respectively. The vector and Fermi multiplets from the D0-D0 strings are in the antisymmetric representation of the gauge group O($k$). The hyper- and twisted hypermultiplets are in the symmetric representation of O($k$). The D0-D4 (D0-D8) strings are in the bifundamental representation of the gauge group O($k$) and flavor group Sp(1) (SO($2N_f$)).

Consider a ${\cal N}=2$ subalgebra with supercharges $Q$ and $Q^\dagger$ inside the ${\cal N}=4$ supersymmetry algebra. The Witten index is defined as
\ie
{\cal Z}^{k}_{\text{\tiny D0-D4-D8/O8}}(t,u,w,m_i)=\Tr_{{\cal H}_{\text{\tiny QM}}}\left[(-1)^F e^{-\beta\{Q^\dagger,Q\}} t^{2(J_++J_R)}u^{2J_-}v^{2J_R'}w^{2\Pi}e^{-\sum F_i m_i}\right],
\fe
where $J_\pm$, $J_R$, $J_R'$ and $\Pi$ are the Cartan generators of the SU(2)$_\pm$, SU(2)$^R_+$, SU(2)$^R_-$ and the Sp(1) flavor symmetry. We give a very brief description of how this Witten index is computed, following \cite{Hwang:2014uwa,Cordova:2014oxa}, by applying supersymmetric localization. The index is invariant under continuous deformations that preserve the supercharges $Q$ and $Q^\dagger$.  One can consider the free field limit, and the path integral over nonzero modes reduces to the product of one-loop determinants, which depend on the fixed background of bosonic zero modes. One then integrates over the zero modes exactly.  

The 1d gauge field is non-dynamical, but its holonomy on S$^1$ is a bosonic zero mode, which combines with the zero mode of the scalar in the vector multiplet to form a complex variable $\phi$ taking values in the maximal torus of the complexified gauge group. There are fermionic zero modes coming from the fermions (gauginos) in the vector multiplet, which are absorbed by the Yukawa coupling terms in the action involving the gauginos, the scalars, and the fermions in the charged matter multiplets (as in (2.21) of \cite{Hwang:2014uwa}). This contributes additional terms to the integrand as the free correlators of the scalars and the fermions, which in turn combine with the previous one-loop determinants to a total derivative of ${\partial/ \partial\bar\phi}$. The zero mode integral becomes a contour integral over $\phi$. The contour can be determined by a careful regularization of the divergences on the complex $\phi$-plane \cite{Benini:2013xpa,Benini:2013nda}, by reintroducing the auxiliary field $D$ of the vector multiplet.

The gauge group O($k$) has two disjoint components.  The group element in one component, denoted by O$(k)_+$, has determinant $+1$, and in the other component, denoted by O$(k)_-$, has determinant $-1$.  
The index is a sum of the $\phi$-contour integrals in each of the components,
\ie\label{eqn:PhiIntegral}
&{\cal Z}^k_{\text{\tiny D0-D4-D8/O8}} = {1\over 2}({\cal Z}^k_+ +{\cal Z}^k_-),
\\
&{\cal Z}^k_\pm = {1\over |W|}\oint [d\phi]{\cal Z}^{\pm,k}_{\text{\tiny D0-D0}}{\cal Z}^{\pm,k}_{\text{\tiny D0-D4}}{\cal Z}^{\pm,k}_{\text{\tiny D0-D8}},
\fe
where $|W|$ is the Weyl factor, i.e., the order of the Weyl group of O($k$), and the integrands are given by the one-loop determinants of the fields listed in \tabref{table:D0-D4-D8/O8}. They are computed in \cite{Kim:2012gu,Hwang:2014uwa}, and we summarize them in \appref{app:integrand}.

For $k=1$ there is no integral, and the integrand directly gives the Witten index. For $k=2$ and $3$, the integrals are one-dimensional. The contour prescription is such that the integrals pick up the residues of the poles coming from the terms in the denominator of the integrand \eqref{eqn:Zplus}-\eqref{eqn:D0-D8integrand} with $+$ sign in front of the $\phi$. For $k\ge 4$, the integrals are multi-dimensional, and the precise contour prescriptions are provided in \cite{Hwang:2014uwa} in terms of Jeffrey-Kirwan residues. (See \appref{app:Technical} for more details about which poles contribute to the integral in the cases $k=2$ and $k=4$.)

We can combine the Witten indices for theories with different $k$ into a generating function
\be
{\cal Z}_{\text{\tiny D0-D4-D8/O8}}(t,u,v,w,m_i,q)=1+\sum_{k=1}^\infty q^k{\cal Z}^{k}_{\text{\tiny D0-D4-D8/O8}}(t,u,v,w,m_i).
\ee
The Nekrasov instanton partition function can be expressed as a ratio of the generating functions \cite{Hwang:2014uwa},
\ie\label{InstantonFromQM}
&{\cal Z}_{\text{\tiny inst}}(t,u,w,m_i,q)={{\cal Z}_{\text{\tiny D0-D4-D8/O8}}(t,u,v,w,m_i,q)\over {\cal Z}_{\text{\tiny D0-D8/O8}}(t,u,v,m_i,q)},
\fe
where ${\cal Z}_{\text{\tiny D0-D8/O8}}(t,u,v,m_i,q)$ is the generating function of the Witten indices of the system with only D0-branes and $N_f$ D8-branes coincident with an O8-plane. It can be obtained by decoupling the D4-brane from our original system,
\ie
{\cal Z}_{\text{\tiny D0-D8/O8}}(t,u,v,m_i,q)=\lim_{w\to 0}{\cal Z}_{\text{\tiny D0-D4-D8/O8}}(t,u,v,w,m_i,q).
\fe
Notice that the Nekrasov instanton partition function on the left hand side of \eqref{InstantonFromQM} is independent of the fugacity $v$ associated with the Cartan of SU(2)$^R_-$. As shown in \cite{Hwang:2014uwa}, the $v$-dependences of the Witten indices in the numerator and the denominator on the right hand side of \eqref{InstantonFromQM} cancel each other.

\section{Line and Ray Operators}
\label{sec:LRO}

The $E_n$ theories possess BPS line operators that preserve half the supersymmetries. They can be realized in the type I' brane construction by probing the D$4$ brane with a fundamental string along directions $9$ and, say, $0$, in analogy with the way a BPS Wilson line was introduced into the low-energy \SUSY{4} SYM on D$3$-branes in \cite{Rey:1998ik,Maldacena:1998im}. The configuration preserves an $\SO(4)\subset\SO(4,1)$ rotation group, as well as those SUSY parameters that satisfy
\be\label{eqn:F1SUSY}
\epsilon_L=\Gamma^{1234}\epsilon_L = \Gamma^{5678}\epsilon_L\,,
\qquad
\epsilon_R = \Gamma^9\epsilon_L\,.
\ee
The position of the line operator can be fixed at $x_1=\cdots=x_4=0$ by introducing an additional D$4$-brane (which we denote by D$4$') at $x_9=L>0$ and requiring the fundamental string to end on it. This D$4$' brane does not break any additional SUSY.
Note that the orientation of the string and D$4$'-brane must be correlated in order to preserve SUSY. Reversing the orientation of the string changes the subspace of $(\epsilon_L,\epsilon_R)$ that are preserved by inserting a $(-)$ in front of $\Gamma^{1234}$ in \eqref{eqn:F1SUSY}.
The directions of the various branes so far are summarized in the first four rows of Table \ref{table:BraneDirections}.
\begin{table}\centering
\begin{tabular}{|c|cccccccccc|l|} \hline
   &  0 &  1 &  2 &  3 &  4 &  5 &  6 &  7 &  8 &  9 & Comments \\
\hline\hline
D8/O8 &\xTB&\xTB&\xTB&\xTB&\xTB&\xTB&\xTB&\xTB&\xTB&    & $N_f\equiv(n-1)$ D8's at $x_9=0$\\
\hline
D4 &\xTB&\xTB&\xTB&\xTB&\xTB&    &    &    &    &    & \\
\hline
F1 &\xTB&   &   &   &   &    &    &    &    &\xTB& $x_0\ge 0$ \\
\hline
D$4'$ &\xTB&    &    &    &    &\xTB&\xTB&\xTB&\xTB&   & $x_9=L>0$\\
\hline
D2 &   &   &   &   &   &    &    &\xTB &\xTB&\xTB& $x_0=0$ \\
\hline
\end{tabular}
\caption{The directions of the various branes.}
\label{table:BraneDirections}
\end{table}
As we reviewed in \secref{sec:Rev5dEnIndex}, the $E_n$ SCFT can be deformed by an operator of dimension $4$ to a theory that flows in the IR to a weakly coupled Sp($1$) gauge theory with flavor group $\SO(2N_f)$, by lowering the string coupling constant $\gst$ to a finite value at the location of the D$8$/O$8$. Denoting by $\SpA_0$ the time component of the gauge field and by $\SpPhi$ its scalar superpartner (i.e., the scalar component of the vector multiplet), the line operator reduces to a supersymmetric Wilson line 
$$
P\exp\bigl[i\int_{-\infty}^\infty (\SpA_0+\SpPhi)dx^0\bigr],
$$
which can be made gauge invariant by compactifying time on $S^1$ and taking the trace.  The operator preserves half of the supercharges.\footnote{The supersymmetry transformation on $\SpA_0$ and $\SpPhi$ can be found in (2.10) of \cite{Kim:2012gu}:
$\delta A_\mu = i\bar \lambda \gamma_\mu \epsilon,$ 
$\delta\Phi = \bar\lambda\epsilon,$
where the index $A$ for the SU(2)$_R$ is implicit. The combination $\SpA_0+\SpPhi$ is preserved by the transformations that satisfy the condition $i\gamma_0\epsilon+\epsilon=0$, which is the condition imposed by fundamental strings. The 5d spinor and 10d spinor are related by $\epsilon = \epsilon_L$ and $i\gamma^\mu=\Gamma^9\Gamma^\mu$.} The gauge theory also has  hypermultiplets with fields in $(\rep{2},2N_f)$ of $\Sp(1)\times$SO$(2N_f)$. 
Denoting the scalar component of these fields by $q^A_\iF$ (with $\iF=1,\cdots,2N_f$ and $A=1,2$), we can construct gauge invariant operators
\ie\label{eqn:WilsonSegment}
\bar q_{2,\,\jF}(t_f)P\exp\left[i\int^{t_f}_{t_i}(A_0+\SpPhi)dx^0\right]q_\iF^1(t_i),
\fe
which preserve half of the SUSY \eqref{eqn:F1SUSY} that the Wilson line preserves, that is two of the eight supercharges of the 5d theory.\footnote{The supersymmetry transformation of $q^A_\iF$ and $\bar q_{A,\iF}$ can be found in (2.14) of \cite{Kim:2012gu}: $\delta q^A_\iF = \sqrt{2}i\bar\epsilon^A\psi_\iF$ and $\delta \bar q_{A,\iF} = \sqrt{2}i\bar\psi_\iF\epsilon_A$.
By the symplectic-Majorana condition $\bar\epsilon^A = \varepsilon^{AB}(\epsilon^T)_B\gamma^2\gamma^4$, the transformations preserving $q^1_\iF$ and $\bar q_{2,\iF}$ satisfy the condition $\epsilon_2=0$, which is equivalent to $\tfrac{i}{2}(\Gamma^{56}-\Gamma^{78})\epsilon_L=\epsilon_L$ for the 10d spinor.} The fields at each endpoint can be locally modified, for example by replacing $\qk^1_\iF(t_i)$ with $\covD_\mu\qk^1_\iF(t_i)$ ($\covD=\partial-i\SpA$).
Most kinds of insertions will break all of the SUSY, but there are some operators that preserve the same amount of SUSY as \eqref{eqn:WilsonSegment} does.  We are interested in counting the number of such BPS operators, with given spin and R-charge, and in testing whether they can be collected into complete $E_n$ multiplets.
We therefore consider {\it ray operators} of the form
\be\label{eqn:WR}
\WR_\cLO=P\exp\bigl[i\int_0^\infty (\SpA_0+\SpPhi)dx^0\bigr]\cLO(0),
\ee
where $\cLO(0)$ is a local operator at $x_0=0$ in the $\rep{2}$ of $\Sp(1)$.

In order to make the case that \eqref{eqn:WR} descend from ray operators in the $E_n$ SCFT, it is useful to modify the type-I' construction by letting the worldsheet of the string end on a (Euclidean) D$2$-brane at $x^0=0$, extending in directions $7,8,9$, as listed in \tabref{table:BraneDirections}.
Introducing a Euclidean brane, which behaves like an instanton, requires us to switch to Euclidean signature (similarly to Yang-Mills theory for which there are no real instanton solutions in Minkowski signature). We therefore Wick rotate $x^0\rightarrow -i x^{\overline{0}}$.
In Euclidean signature, the Weyl spinor condition is\footnote{There is also a reality condition $\epsilon_L^\star=C\epsilon_R$, where $C$ is the charge conjugation matrix.}
\ie
i\Gamma^{\overline{0}123456789 }\epsilon_R = \epsilon_R ,~~~i\Gamma^{\overline{0}123456789 }\epsilon_L = -\epsilon_L.
\fe
The SUSY generators preserved by a Euclidean D$p$-brane in directions $\overline{0},\dots,p$ satisfy the condition
$\epsilon_R = i\Gamma^{\overline{0} 1 \cdots p}\epsilon_L.$
The generators preserved by a fundamental string in the $9^{th}$ direction satisfy the conditions
$\epsilon_R = i\Gamma^{\overline09 }\epsilon_R$ and $\epsilon_L = - i\Gamma^{\overline09 }\epsilon_L.$
Combining these conditions we obtain the same conditions on the supersymmetry generators as those imposed by the Euclidean D$4$-D$8$/O$8$ configuration, which is given by the same equations as \eqref{eqn:F1SUSY}.

The D$2$-brane provides an anchor for the F$1$ to end on.
It breaks half of the remaining supersymmetries, preserving only those SUSY parameters that satisfy
\be\label{eqn:F1D2SUSY}
\epsilon_L=\Gamma^{1234}\epsilon_L = -i\Gamma^{56}\epsilon_L
=i\Gamma^{78}\epsilon_L\,,
\qquad
\epsilon_R = \Gamma^9\epsilon_L\,.
\ee
The F$1$-D$2$-D$4$-D$8$/O$8$ configuration thus preserves only two linearly independent supercharges, and we can take one of them to coincide with $Q=Q_2^1$ of \eqref{BPSbound}.
Indeed, a generator of a Cartan subalgebra of the SU(2)$_R$ that acts on the index $A$ of $Q_m^A$ can be identified with $\tfrac{i}{2}(\Gamma^{78}-\Gamma^{56})$, and the index $A=1$ can be defined to label the eigenspace of $\tfrac{i}{2}(\Gamma^{78}-\Gamma^{56})$ with eigenvalue $+1$. Similarly, $\tfrac{i}{2}(\Gamma^{12}+\Gamma^{34})$ can be identified with $J_{-}$, which was defined below \eqref{SCA}, and the index $m=2$ is one of the two indices that generate an eigenspace of $J_{-}$ with eigenvalue (i.e., spin) zero.
The insertion of an F$1$ ending on a Euclidean D$2$-brane thus preserves two out of the eight supercharges $Q_m^A$ of the 5d SCFT that describes the low-energy of the D$4$-brane in the background of the D$8$/O$8$ system. 

We will be interested in extending the superconformal index \eqref{SuperconformalIndex} to ray operators of the form \eqref{eqn:WR}, and the operators that will contribute to our index need to preserve only one supercharge ($Q_2^1$). Such operators can appear on the F1 $\cap$ D4 intersection by coupling to operators on the two dimensional D2 $\cap$ (D8/O8) intersection.
More precisely, the low-energy action of the brane configuration includes factors schematically of the form
\begin{eqnarray}
\lefteqn{
S({\rm F1}-{\rm D2}-{\rm D4}-{\rm (D8/O8)})
\sim
}\nonumber\\ &&
S_{9d}({\rm D8/O8})
+S_{5d}({\rm D4})
+S_{3d}({\rm D2})
+S_{2d}({\rm F1})
\nonumber\\ &&
+S_{2d}({\rm D2}\cap {\rm (D8/O8)})
+S_{1d}({\rm F1}\cap {\rm D2})
+S_{1d}({\rm F1}\cap {\rm D4})
+S_{0d}({\rm D2}\cap {\rm D4}\cap{\rm (D8/O8)})
\,.
\nonumber
\end{eqnarray}
The 2d intersection D2 $\cap$ (D8/O8) supports an $E_n$ chiral current algebra at level $k=1$ [denoted $(\widehat{E_n})_1$], and the exponent of the terms $-S_{1d}({\rm F1}\cap {\rm D4})$ and $-S_{0d}$ is expected to be a sum of products of a ray operator of the $E_n$ SCFT and a local operator of the ${\rm F1}\cap {\rm (D8/O8)}$ theory, of the form $\sum_\alpha\WR_\alpha\OCCA_{\alpha}(0)$,
where $\OCCA_{\alpha}$ is some local operator in a representation of  $(\widehat{E_n})_1$. This suggests a connection between the multiplicities of ray operators and representations of $(\widehat{E_n})_1$, which we will explore elsewhere \cite{toAppear}.

\subsection{The states corresponding to a ray operator}
\label{subsec:StatesR}

The state-operator correspondence of 5d SCFTs assigns to a local operator a gauge invariant state in the Hilbert space of the theory on $S^4$ via a conformal transformation that acts on the radial coordinate as $r\rightarrow \tau=\log r$.
This state-operator correspondence converts the ray operator to a state in the Hilbert space of the theory on S$^4$ with an impurity at one point of S$^4$, which we shall refer to as the {\it South Pole} (\SP).
After flowing to the $\Sp(1)$ gauge theory, the impurity is replaced with an external quark at \SP\ in the fundamental representation of $\Sp(1)$.
Let $\gGaugeAll$ be the (infinite dimensional) group of $\Sp(1)$ gauge transformations on S$^4$, and let $\gGauge\subset\gGaugeAll$ be the group of gauge transformations that are trivial at \SP.
Then $\gGaugeAll/\gGauge\cong\Sp(1)$ and the states that correspond to ray operators are those that are invariant under $\gGauge$ but are doublets of $\gGaugeAll/\gGauge$.
We will ``count'' them, or rather calculate their supersymmetric index, by inserting a Wilson loop at \SP\ into the partition function of the $\Sp(1)$ gauge theory on S$^4$$\times$S$^1$, as will be explained in detail in \secref{sec:IndexRO}.

As we argued above, the ``impurity'' at \SP\ preserves the SUSY generators with parameters restricted by \eqref{eqn:F1D2SUSY}. In the notation of \secref{subsec:Index}, these are the generators $Q_m^A$ with $A=1$ and $m=1,2$ (and the generators with $m=3,4$ or $A=2$ are generally not preserved). The impurity preserves the inversion $\tau\to -\tau$; hence, also preserves the superconformal generators $S^m_A$ with $A=1$ and $m=1,2$. The bosonic subalgebra that preserves a ray is generated by the dilatation operator $D$, the generators $M_1{}^1=-M_2{}^2$, $M_1{}^2$, $M_2{}^1$, $M_3{}^3=-M_4{}^4$, $M_3{}^4$, $M_4{}^3$ of the rotation subgroup SU(2)${}_+\times$SU(2)${}_{-}\cong$ Spin(4) $\subset$ Spin(5), and the R-symmetry generators $R_1{}^1$. The above bosonic and fermionic generators form a closed subalgebra of $F(4)$. (See \cite{Cordova:2016uwk} for a discussion of the subalgebra preserved by a ray in 4d superconformal theories.)

In order to extend the discussion of \secref{subsec:Index} to states on $S^4$ with impurity at \SP, we need to establish first which subalgebra of the superconformal algebra $F(4)$ acts on the Hilbert space. Such subalgebra properly contains the subalgebra preserved by a ray.\footnote{This distinction also occurs in the case of local operators, where the origin is only preserved by dilatations and by SO(5)$\times$SU(2)${}_R$, but translations $P_\mu$ and conformal transformations $K_\mu$ are good operators on the Hilbert space, defined as the space of states on $S^4$.} It also contains the translation operator $P_{\overline{0}}$ and conformal generator $K_{\overline{0}}$.
Note that neither of these preserve the ray --- $P_{\overline{0}}$ does not preserve the origin, while $K_{\overline{0}}$ does not preserve the endpoint at infinity.
Nevertheless, if we define the Hilbert space at, say, $r=1$, both generators preserve the location of the impurity on S$^4$. We can now obtain the full set of $Q_m^A$ and $S_A^m$ with $m=1,2$ and $A=1,2$ by starting with $Q=Q_2^1$ and its hermitian conjugate $S=S_1^2$ and successively calculating commutators with the bosonic generators $K_{\overline{0}}$, $P_{\overline{0}}$, $M_1{}^2$ and $M_2{}^1$. 

The BPS bound \eqref{BPSbound} is therefore valid for ray operators, too. 
Note, however, that $Q_3^A$ and $Q_4^A$ are not preserved by the line impurity at \SP, and there is no way to get them from commutators of $Q$ with SU(2)${}_{+}\times$SU(2)${}_-$ generators. We therefore cannot assume \eqref{eqn:BPSboundQ4S4}.
Nevertheless, the parts of the discussion at the end of \secref{subsec:Index} that do not rely on \eqref{eqn:BPSboundQ4S4} are still valid. In particular, we can define an index similarly to \eqref{SuperconformalIndex}, and it receives contributions only from nontrivial elements of the $Q$-cohomology. Moreover, states that contribute to the index have maximal $J_{+}+J_R$ in their SU(2)${}_{+}\times$SU(2)${}_R$ multiplet.
It follows that no state that contributes to the index can have $J_{+}+J_R=0$, because if it did it would be a singlet of SU(2)${}_{+}\times$SU(2)${}_R$ and thus would be annihilated by $R_1{}^2$ and $M_1{}^2$, and therefore also by the commutator $[M_1{}^2,[R_1{}^2,Q_2{}^1]]\propto Q_1{}^2$. But $\{Q_1{}^2,Q_2{}^1\}\propto P_{\overline{0}}$, which does not preserve a ray operator. This observation will become relevant in \secref{subsec:IndexRays} when we preserve our result for the index of the $E_8$ theory.

The calculation of the index of ray operators that will follow makes the $\SO(2n-2)\times$U(1) $\subset E_n$ global symmetry explicit, but in order to properly combine the $\SO(2n-2)\times$U(1) characters into $E_n$ characters it is important to first explain a shift in the U(1) charge.

\subsection{Shifted instanton number}
\label{subsec:shiftI}
Denote the U(1) charge by $\qQ$.
On local operators that correspond to gauge invariant states on S$^4$, the U(1) charge is simply the integer instanton number
$$
\qQ = \frac{1}{8\pi^2}\int_{\text{S}^4}\tr(F\wedge F) = \kInst.
$$
However, on states that correspond to a ray operator, the $U(1)$ charge receives an anomalous contribution and reads
\be\label{eqn:qQshift}
\qQ = \kInst+\frac{2}{N_f-8}\,,\qquad (N_f=2,\dots,7).
\ee
The correction $2/(N_f-8)$, which is fractional for $N_f< 6$, will be borne out by the index that we will present in \secref{sec:IndexRO}. Below, we will review the physical origin of this shift.
Our discussion is similar to the arguments presented in \cite{Brodie:2000ns,Bergman:2001qg}.

The shift \eqref{eqn:qQshift} is easy to explain on the Coulomb branch of the $E_n$ theory by using the D$4$-D$8$/O$8$ brane realization. The space $x^9>0$ is described by massive type-IIA supergravity at low-energy with mass parameter proportional to $\mFlux\defineas 8-N_f$ \cite{Polchinski:1995df}. In \appref{app:miia}, we review the D8-brane solution in the massive type-IIA supergravity, and the D-branes worldvolume actions in that background. The Coulomb branch corresponds to the D$4$-brane moving away from the D$8$/O$8$ plane in the positive $x^9$ direction. The low-energy description is a free U(1) vector multiplet. Denote the vector field by $\UoneA$, the field strength by $\UoneF=d\UoneA$, and scalar component by $\UonePhi$.  The scalar component has a nonzero VEV $\PhiVEV=\langle\UonePhi\rangle>0$ (proportional to the $x^9$ coordinate of the D$4$-brane). 

We can understand the shift \eqref{eqn:qQshift} in the U(1) charge after reviewing the peculiar interaction terms that are part of the low-energy description of a D-brane in massive type-IIA supergravity \cite{Romans:1985tz}.
As we review in \appref{app:miia}, the super Yang-Mills effective action on a D$p$-brane includes an additional Chern-Simons term proportional to $\mFlux \UoneA\wedge \UoneF^{p/2}$ \cite{Green:1996bh}. It implies a few modifications to the conservation of string number. For $p=0$, we find that a net number of $\mFlux$ fundamental strings must emanate from any D$0$-brane.
As usual, a D$0$-brane can be absorbed by a D$4$-brane and convert into one unit of instanton charge. In that case, the $\mFlux$ strings that are attached to the D$0$-brane can convert to $\mFlux$ units of electric flux. Indeed, the low-energy description of the D$4$-brane, which is the low-energy effective action of the 5d $E_n$ theory on the Coulomb branch \cite{Intriligator:1997pq}, contains an effective Chern-Simons interaction term proportional to $\mFlux\UoneA\wedge\UoneF\wedge\UoneF$, and  can be written as
\be\label{eqn:ICoulomb}
I_{\text{Coulomb}} =-\int\Bigl\lbrack
\frac{1}{8\pi^2}\mFlux\PhiVEV\UoneF\wedge{}^\star\UoneF
+\frac{1}{24\pi^2}\mFlux\UoneA\wedge\UoneF\wedge\UoneF
-\UoneA\wedge{}^\star\UoneJ
\Bigr\rbrack\,,
\ee
where $\UoneJ$ is the contribution of the hypermultiplet to the $U(1)$ current.
The $\UoneA_0$ equation of motion can be written as
\be\label{eqn:A0EOM}
{1\over 4\pi^2}\mFlux\PhiVEV d({}^\star\UoneF)
 = \frac{1}{8\pi^2}\mFlux\UoneF\wedge\UoneF
-{}^\star\UoneJ
\,,
\ee
Integrating \eqref{eqn:A0EOM} over 4d space shows that $\mFlux/2$ units of electric flux accompany one unit of instanton charge.

It follows that we can measure the U(1) charge $\qQ$ in two equivalent ways, by either (i) integrating ${1\over 4\pi^2}\UoneF\wedge\UoneF$ over all of space\footnote{In our convention, the SU(2) instanton number is related to the U(1) instanton number by
\ie
\kInst = \frac{1}{8\pi^2}\int_{\text{S}^4}\tr(F\wedge F) =  {1\over 4\pi^2}\int_{\text{S}^4} \UoneF\wedge\UoneF,
\fe
where we have used $F=\UoneF \sigma_3$.}, or (ii) measuring the electric flux ${\mFlux\PhiVEV\over 4\pi^2} \int{}^\star\UoneF$ at infinity and dividing by $\mFlux/2$.
In a general situation, however, there could be a net number $\nStrings$ of open fundamental strings attached to the D$4$-brane and extending into the bulk $x^9$ direction. Their endpoints on the D$4$-brane behave as external charges, which contribute $\nStrings\delta^4(x)$ to ${}^\star\UoneJ$, and then methods (i) and (ii) above will give a different answer for $\qQ$. The answers differ by $2\nStrings/\mFlux$. To determine which method is correct, we note that an instanton can evaporate into the $x^9$-bulk as a D$0$-brane, which would then carry with it $\mFlux$ open strings. Such a process reduces the instanton number $\kInst$ by $2$ (the factor of $2$ is the effect of the orientifold), and at the same time reduces $\nStrings$ by $\mFlux$. The possibility of such a process demonstrates that instanton number alone is not conserved, and method (ii) is the correct one. This is consistent with the shift of $-2/\mFlux$ in \eqref{eqn:qQshift}.

\subsection{The center of $E_n$}
\label{subsec:shiftI}

We will now discuss the action of the center $\Zcenter_n$ of the enhanced $E_n$ flavor symmetry.
For $n\ge 3$, we will use the convention that $E_n$ is simply connected.
For $n=3,\dots,7$, $E_n$ then has a nontrivial center given by $\Zcenter_n\cong\Z_{9-n}=\Z_{8-N_f}$. For $n=2$ we have $E_2=$ SU(2)$\times$U(1) which has $\Zcenter_2\cong\Z_2\times$U(1) as center.
So far, when discussing the ``SO$(2N_f)\times$U(1) subgroup'' of $E_n$, we have not been precise about the global structure, which we will now rectify.

Local operators of the $E_n$ SCFT are neutral under $\Z_{9-n}$.
Indeed, the only $E_n$ representations of local operators found in \cite{Hwang:2014uwa} have weights belonging to the root lattice.
In contrast, ray operators are charged under $\Z_{9-n}$.
Let $\latQrt{n}$ be the root lattice of $E_n$, and let $\latQwt{n}$ be the weight lattice. We will find in \secref{subsec:IndexRays} representations whose weights project to a nontrivial element of $\latQwt{n}/\latQrt{n}\cong\Z_{9-n}$. For $3\le n\le 7$, this $\Z_{9-n}$ can be identified with the Pontryagin dual of the center $\Zcenter_n$.
In other words, ray operators carry nontrivial $\Z_{9-n}$ charge.

Consider $N_f=6$, for example. $E_7$ has a subgroup [$\Spin$(12)$\times$SU(2)]/$\Z_2$, where the $\Z_2$ identification means the following.
 Denote by $\rep{32}$ one of the two chiral spinor representations of $\Spin(12)$, and by $\rep{32}'$ the other one. $\Spin(12)$ has a center $\Z_2'\times\Z_2''$, where the generator of $\Z_2'$ is defined to be $(-1)$ in $\rep{12}$ and $\rep{32}'$, and the generator of $\Z_2''$ is defined to be $(-1)$ in $\rep{12}$ and $(-1)$ in $\rep{32}$; the $\Z_2'\times\Z_2''$ charges in other representations of $\Spin(12)$ are defined by requiring additivity mod $2$ under tensor products.  Then, when decomposing representations of $E_7$ into irreducible representations of $\Spin(12)\times$SU(2), half-integer SU(2) spins will always be paired with representations of $\Spin(12)$ that are odd under $\Z_2'$, while even SU(2) spin will be paired with zero $\Z_2'$ charge. So, for example, $(\rep{12},\rep{2})$, $(\rep{32}',\rep{2})$, and $(\rep{32},\rep{1})$ are allowed, but neither $(\rep{12},\rep{1})$ nor $(\rep{1},\rep{2})$ nor $(\rep{32},\rep{2})$ can appear. Taking the U(1) $\subset$ SU(2) subgroup, we find the subgroup $[\Spin(12)\times$U(1)]/$\Z_2\subset E_7$ under which the fundamental representation $\rep{56}$ and adjoint $\rep{133}$ decompose as
$$
\rep{56} = \rep{12}_{1}+\rep{12}_{-1}+\rep{32}_0,
\qquad
\rep{133} = \rep{1}_{2}+\rep{1}_0+\rep{66}_0+\rep{1}_{-2}
+\brep{32}'_{1}+\rep{32}'_{-1}.
$$
In our conventions, one unit of instanton number ($k=1$) corresponds to one unit of the above U(1) charge.
The generator of $\Z_2$ that appears in $[\Spin(12)\times U(1)]/\Z_2$ is therefore identified with $(-1)^k$ times the generator of $\Z_2'\subset \Z_2'\times\Z_2''\subset\Spin(12)$. The center of $E_7$ is identified with the other factor, $\Z_2''\subset\Spin(12)$. We will see that ray operators are odd under $\Z_2''$.

For $E_6$ ($N_f=5$), the center is $\Z_3$.
The representations of ray operators that we will find are $\rep{27}$, $\rep{1728}$, etc., and the ray operator has one unit of $\Z_3$ charge.
$E_6$ has a subgroup $[\Spin(10)\times$U(1)]$/\Z_4\subset E_6$. The $\Z_4$ identification means the following. The center of $\Spin(10)$ is $\Z_4$, and a representation of $\Spin(10)$ can be assigned a $\Z_4$ charge by the rules that: (i) the $\Z_4$ charge is additive under tensor products; (ii) the left-chirality spinors $\rep{16}$ are assigned charge $1$ mod $4$.
Thus the fundamental $\rep{10}$ is assigned $2$ mod $4$, the adjoint is assigned $0$, and the right-chirality $\brep{16}$ is assigned charge $3$ mod $4$.
Then, when decomposing any representation of $E_6$ under $\Spin(10)\times$U(1), a $\Spin(10)$ representation with $\Z_4$ charge $\gamma$ will always carry U(1) charge that is $\gamma$ mod $4$.
For example, $\rep{27}$ of $E_6$ decomposes under SO(10)$\times$U(1) as
$$
\rep{27}\rightarrow \rep{10}_{-2}+\rep{16}_{1}+\rep{1}_{4}.
$$
In our conventions, one unit of instanton number ($k=1$) corresponds to $3$ units of U(1) charge. Thus, any of the states of $\rep{10}$ carry $-2/3$ instanton number, the states of $\rep{16}$ carry $1/3$ instanton number and $\rep{1}$ carries $4/3$.
The center $\Z_3$ is generated by the projection to  $[\Spin(10)\times$U(1)]/$\Z_4$ of the element $(1,e^{2\pi i/3})\in\Spin(10)\times$U(1).
Thus, for the case $N_f=5$ we see that ray operators carry instanton charge in $\frac{1}{3}+\Z$ and are charged one unit under $\Z_3$.

For $N_f=4$, we have $E_5=\Spin(10)$ and the center of $\Spin(2N_f)=\Spin(8)$ is $\Z_2'\times\Z_2''$ with $\Z_2'$ nontrivial in the vector representation $\rep{8}_v$ and the spinor representation $\rep{8}_s$, and trivial in the spinor representation $\rep{8}_c$ of opposite chirality, while $\Z_2''$ is trivial in $\rep{8}_v$ and nontrivial in both spinor representations.
Then $E_5=\Spin(10)=[\Spin(8)\times$U(1)$]/\Z_2$, where one unit of U(1) charge is identified with instanton number $k$ and the generator of the last $\Z_2$ is identified with $(-1)^k$ times the generator of $\Z_2'$. Thus, in decomposing a representation of $\Spin(10)$ into representations of $\Spin(8)\times$U(1), even U(1) charge is paired with representations of $\Spin(8)$ that appear in tensor products of $\rep{8}_c$, while odd U(1) charge is paired with $\rep{8}_v$ or representations that appear in tensor products of one factor of $\rep{8}_v$ and an arbitrary number of $\rep{8}_c$. 
The center $\Z_4\subset\Spin(10)$ is generated by $k$ ${\pmod{4}}$ plus twice the $\Z_2''$ charge. So, for example, the fundamental representation $\rep{10}$ of $\Spin(10)$ decomposes under $\Spin(8)\times$U(1) as\footnote{Note that $\rep{8}_c$ and not $\rep{8}_v$ appears in the decomposition above, and the embedding of $\Spin(8)\times$U(1) is not the standard one, but related to it by triality.}
$$
\rep{10}=(\rep{8}_c)_0+\rep{1}_2+\rep{1}_{-2}\,.
$$
The states of $\rep{10}$ have $\Z_4$ charge $2\pmod{4}$, while
$$
\rep{16}=(\rep{8}_v)_1+(\rep{8}_s)_{-1}
$$
has charge $1\pmod{4}$, and 
$$
\brep{16}=(\rep{8}_v)_{-1}+(\rep{8}_s)_{1}
$$
has charge $3\pmod{4}$.
The trivial representation $\rep{1}$ and the adjoint $\rep{45}$ of $\Spin(10)$ have $\Z_4$ charge $0\pmod{4}$.
We will find that ray operators have $\Z_4$ charge $1\pmod{4}$. (Whether it is $1$ or $-1$ is a matter of convention.)

For $N_f=3$, we have $E_4=$ SU(5) with center $\Z_5$ and subgroup $[\Spin(6)\times$U(1)$]/\Z_4\subset$ SU(5). Here a generator of $\Z_4$ can be taken as a generator of the center $\Z_4\subset\Spin(6)=$ SU(4) times $i\in$ U(1). A generator of the center of SU(5) can be taken as $e^{2\pi i/5}\in$ U(1) [times the identity in $\Spin(6)$]. We will find that ray operators have one unit of charge under $\Z_5$, meaning that only representations that have Young diagrams with number of boxes equal to $3\pmod{5}$ can appear.

For $N_f=2$, we have $E_3=$ SU(3)$\times$SU(2), and $\Spin(2N_f)\times$U(1) $=$ SU(2)$'\times$SU(2)$''\times$U(1) is related to $E_3=$ SU(3)$\times$SU(2) by identifying the SU(2) factor with SU(2)$'$, and noting the subgroup [SU(2)$''\times$ U(1)]/$\Z_2\subset$ SU(3). Again, the $\Z_2$ identification means that in decomposing representations of SU(3) under SU(2)$''\times$U(1), odd SU(2)$''$ spin is paired with odd U(1) charge, and vice versa. The center of $E_3$ is $\Z_3\times\Z_2'$. Referring to SU(2)$'\times$[SU(2)$''\times$U$(1)]/\Z_2$ $\subset E_3$, the generator of $\Z_3\subset E_3$ can be identified with $e^{2\pi i/3}\in$ U(1), and $\Z_2'\subset E_3$ is identified with the center of SU(2)$'$.
We will see that ray operators have charge $2\pmod{3}$ under $\Z_3$, and they are odd under $\Z_2'$.

For $N_f=1$, we have $E_2=$ SU(2)$\times$U(1), and the center is $\Z_2\times$U(1).
We will find that ray operators carry fractional U(1) charge in $\tfrac{4}{7}+\Z$ and their $\Z_2$ charge is correlated with their U(1) charge. More details on the definition of $E_2$ and the embedding of SO$(2N_f)=$ SO(2) in it are reviewed in \appref{app:E2}.


\section{The index of line and ray operators}
\label{sec:IndexRO}

\subsection{Wilson ray indices}
\label{subsec:IndexROwr}
The spectrum of line or ray operators can be studied by computing the line/ray operator indices analogous to the superconformal indices.  Let us first discuss Wilson line operators. Consider a line operator supported on a line $\R^1\subset\R^5$, which without loss of generality we can choose to pass through the origin of $\R^{5}$. By a conformal map to S$^4\times\R^1$, the line $\R^1\subset\R^5$ is mapped to two lines at antipodal points $\SPole,\NPole\in$S$^4$ and along the $\R^1$ factor of S$^4\times\R^1$, with opposite orientations for the two lines $\{\SPole\}\times\R^1$ and $\{\NPole\}\times\R^1$. Similarly to the superconformal index, the Wilson line index can be computed by a path integral on S$^4\times$S$^1$. Since the path integral localizes on solutions of constant holonomy $U=e^{i\int_{{\rm S}^1} A_\mu dx^\mu}$ along the ``thermal'' circle S$^1$, the Wilson line operators simply reduce to the Sp(1) characters
\ie\label{eqn:chiRw}
\chi_{\bf R}(w)=\tr_{\bf R} U.
\fe
Since the Wilson lines at the antipodal points have opposite orientations, they correspond to characters of conjugate representations. For Sp(1), conjugate representations are equivalent, but since the construction generalizes to Sp($N$) with any $N$, we will retain the distinction between a representation ${\bf R}$ and its conjugate ${\bf \bar{R}}$. This discussion suggests that the Wilson line index can be calculated by inserting a pair of characters of opposite representations into the integral formula \eqref{SCI} of the superconformal index \cite{Gang:2012yr},
\be\label{eqn:WLnaive}
{\cal I}^{\text{\tiny Wilson line}}_{\bf R}(t,u,m_i,q)
\stackrel{?}{=}\int _{\rm Sp(1)}\chi_{\bf R}(w)\chi_{\bf \bar{R}}(w) {\cal Z}_{\text{\tiny1-loop}}(t,u,w,m_i) |{\cal Z}_{\text{\tiny inst}}(t,u,w,m_i,q)|^2dU.
\ee
The question mark over the equality sign indicates that \eqref{eqn:WLnaive} is not the complete answer, as we will discuss below, in the context of ray operators.

The ray operator indices can be studied in a similar way. Consider a ray operator located on a half line ${\mathbb R}^+$, whose end point is chosen to be the origin of the $\R^5$ spacetime. Under the conformal map, the ray operator is mapped to a line operator along the $\R^1$ factor of S$^4\times\R^1$ and located at a point on S$^4$. The Wilson ray operator index therefore appears to be given by the formula
\be\label{eqn:WRnaive}
{\cal I}^{\text{\tiny Wilson ray}}_{\bf R}(t,u,m_i,q)
\stackrel{?}{=}\int _{\rm Sp(1)}\chi_{\bf R}(w){\cal Z}_{\text{\tiny1-loop}}(t,u,w,m_i) |{\cal Z}_{\text{\tiny inst}}(t,u,w,m_i,q)|^2dU.
\ee
For example, let us consider the indices of Wilson rays in the fundamental representation. For $0\leq N_f\leq 7$, the indices in the $t$-expansion are given by

\ie\label{lineindex}
&N_f=0:\hspace{0.2cm}{\cal I}^{\text{\tiny Wilson ray}}_{\bf 2}\stackrel{?}{=}0,
\\
&N_f=1:\hspace{0.2cm}{\cal I}^{\text{\tiny Wilson ray}}_{\bf 2}\stackrel{?}{=}2t+\bigg(\frac{1}{q}+q\bigg)t^3+O(t^4),
\\
&N_f=2:\hspace{0.2cm}{\cal I}^{\text{\tiny Wilson ray}}_{\bf 2}\stackrel{?}{=}4t+\bigg(\frac{6}{q}+16+6q\bigg)t^3+O(t^4),
\\
&N_f=3:\hspace{0.2cm}{\cal I}^{\text{\tiny Wilson ray}}_{\bf 2}\stackrel{?}{=}6t+\bigg(\frac{20}{q}+64+20q\bigg)t^3+O(t^4),
\\
&N_f=4:\hspace{0.2cm}{\cal I}^{\text{\tiny Wilson ray}}_{\bf 2}\stackrel{?}{=}8t+\bigg(\frac{56}{q}+160+56q\bigg)t^3+O(t^4),
\\
&N_f=5:\hspace{0.2cm}{\cal I}^{\text{\tiny Wilson ray}}_{\bf 2}\stackrel{?}{=}10t+\bigg(\frac{144}{q}+320+144q\bigg)t^3+O(t^4),
\\
&N_f=6:\hspace{0.2cm}{\cal I}^{\text{\tiny Wilson ray}}_{\bf 2}\stackrel{?}{=}12t+\bigg(\frac{12}{q^2}+\frac{352}{q}+560+352q+12q^2\bigg)t^3+O(t^4),
\\
&N_f=7:\hspace{0.2cm}{\cal I}^{\text{\tiny Wilson ray}}_{\bf 2}\stackrel{?}{=}14t+\bigg(\frac{195}{q^2}+\frac{832}{q}+896+832q+195q^2\bigg)t^3+O(t^4),
\fe
where we have turned off the fugacities associated to the flavor SO($2N_f$) group.  A few comments on the above formulas are in order. The leading terms of the $t$-expansions in \eqref{lineindex} correspond to the Wilson rays contracted with the scalar $q^1$ of the hypermultiplet [see \eqref{eqn:WilsonSegment}]. As we discussed in \secref{subsec:SCIfromSYM}, the operator $q^1$ represents a nontrivial class of the $Q$-cohomology, has $J_R$ charge $1/2$, and transforms in the $(\rep{2},{\bf 2N_f})$ of $\Sp(1)\times\SO(2N_f)$.  Hence, it contributes a term $2N_f t$ to the Wilson ray index.  The flavor symmetry of the IR Sp(1) SYM combines with the instanton number symmetry U(1) and is enhanced to form the $E_n$ global symmetry of the UV CFT. Under the broken generators of $E_n\to\SO(2N_f)\times$U(1), the Wilson ray operators are transformed to ray operators of nonzero instanton U(1) charge. However, our Wilson ray indices \eqref{lineindex} did not capture those contributions. One would expect the full ray operator indices to exhibit the structure of the $E_n$ symmetry; more precisely, the coefficients of the $t$-expansion must be characters of $E_n$, but this is not the case in \eqref{lineindex}. For example, for $N_f=6$, the representation $\bf 12$ that appears in the leading $t$-expansion of the Wilson ray index should be completed to the representation $\bf 56$ of $E_7$.  This instructs us to look for additional contributions to the Nekrasov instanton partition function ${\cal Z}_{\text{\tiny inst}}$.

The problem with the naive prescription \eqref{eqn:WRnaive} is that it relies on the evaluation of the Wilson loop \eqref{eqn:chiRw} at a point (the south pole) where the gauge field configuration is singular (a zero size instanton).
To resolve the singularity, we have to invoke a string theory construction similar to \tabref{table:BraneDirections}. We then see that in the presence of zero-size instantons (interpreted as D$0$-branes) the fundamental string (F1) can end on either the D$4$-brane directly or on a D$0$-brane. If the F$1$ ends on the D$4$ it induces the Wilson loop term \eqref{eqn:chiRw} in the action, but if F$1$ ends on a D$0$-brane, with $k$ D$0$-branes present, it will manifest itself as an O$(k)$ Wilson loop. The Hilbert space of the F$1$-D$4$-D$8$/O$8$ system thus has several sectors.

Instead of analyzing each string sector separately and adding up the contributions, it is much more convenient to compute a generating function for the partition function with an arbitrary number $\lStrings$ of F$1$ strings. We therefore introduce a new variable $x$, which will play the role of fugacity for the string number, so that the terms of order $x^{\lStrings-1}$ in the generating function will capture the partition function with $\lStrings$ strings present. (The shift by $-1$ will be explained shortly below.) 
So, we propose a formula for the generating function of the ray operator indices,
\ie\label{eqn:ROI}
{\cal I}_{\text{\tiny all}}&(t,u,x,m_i,q)
\\
&=\int _{\rm Sp(1)} {\cal Z}_{\text{\tiny1-loop}}(t,u,w,m_i)  \ZinstLine(t,u,w,x,m_i,q){\cal Z}_{\text{\tiny inst}}(t,u,w^{-1},-m_i,q^{-1})dU,
\fe
where $\ZinstLine$ is the instanton partition function on the background of a a line operator on $\R^1\subset\R^5$ with the Omega background turned on (on the space $\R^4$ transverse to the line operator). 

To compute $\ZinstLine$ we follow a technique developed in \cite{Tong:2014cha,Nekrasov:2015wsu,Kim:2016qqs} and introduce a D$4'$-brane on which F$1$ can end (as in \tabref{table:BraneDirections}). The D$0$-D$4$-D$4'$-D$8$/O$8$ system then automatically allows for a dynamical generation of finite-mass F$1$-strings. The D$4'$ brane supports a U(1) gauge field, and $x$ is more precisely identified as the fugacity for this U(1) charge.
The presence of the D$4$-brane generates nontrivial RR four-form flux that induces a background of $(-1)$ units of U$(1)$-charge\footnote{The D$4$-D$4'$ system is T-dual to a D$0$-D$8$ system, and the effect is similar to the induced charge of $\mFlux$ units discussed in \secref{subsec:shiftI}.}, and so the sector with $\lStrings$ F$1$-strings has U$(1)$ charge $(\lStrings-1)$.

Similarly to Nekrasov's instanton partition function ${\cal Z}_{\text{\tiny inst}}$, the modified partition function $\ZinstLine$ can be computed as a ratio of Witten indices of D0-brane quantum mechanics systems,
\ie\label{eqn:LineZoverZ}
\ZinstLine(t,u,w,x,m_i,q)={{\cal Z}_{\text{\tiny D0-D4-D$4'$-D8/O8}}(t,u,v,w,x,m_i,q)\over {\cal Z}_{\text{\tiny D0-D$4'$-D8/O8}}(t,u,v,x,m_i,q)}.
\fe
The D0-D4-D$4'$-D8/O8 and D0-D$4'$-D8/O8 quantum mechanics systems will be discussed in detail in the next section. The partition function $\ZinstLine$ is expected to be independent of the fugacity $v$ associated with the Cartan subgroup of the SU(2)$_-^R\subset$ Spin(4)$_{5678}$ rotation group, since none of the gauge theory degrees of freedom are charged under it. For $N_f<7$, we checked up to ${\cal O}(t^5)$ order that the $v$-dependences of the numerator and denominator on the right hand side of \eqref{eqn:LineZoverZ} cancel each other. For $N_f=7$, however, the right hand side of \eqref{eqn:LineZoverZ} does depend on the fugacity $v$. 
Thus, in this special case we see that ${\cal Z}_{\text{\tiny D0-D4-D$4'$-D8/O8}}$ does not factorize into a decoupled $E_8$ SCFT contribution and ${\cal Z}_{\text{\tiny D0-D$4'$-D8/O8}}$. We will return to this problem in \secref{subsec:IndexRays}.

Setting aside the problem of $N_f=7$, for now,
the ray operator indices are then extracted from the generating function ${\cal I}_{\text{\tiny all}}$ by expanding in $x$,
\ie\label{eqn:IallExpansion}
 {\cal I}_{\text{\tiny all}}(t,u,w,x,m_i,q) = {1\over x} {\cal I}_{\text{\tiny SCI}}(t,u,w,m_i,q) -  {\cal I}_{\text{\tiny ray}}(t,u,w,m_i,q) + {\cal O}(x).
\fe
The first term of the expansion is the superconformal index ${\cal I}_{\text{\tiny SCI}}$, and the second term ${\cal I}_{\text{\tiny ray}}$ is the ray operator index. The minus sign in front of ${\cal I}_{\text{\tiny ray}}$ is because the single D$4-$D$4'$ string is fermionic.

\subsection{The D0-D4-D4$'$-D8/O8 system}\label{modifiedadhm}

In \cite{Tong:2014cha}, the interaction of instantons and Wilson line operators was studied by introducing an extra D4-brane (referred to as a D$4'$-brane) to the D0-D4 quantum mechanics. The D$4'$-brane has no spatial direction in common with the D4-branes in the D0-D4 system. The directions of the D4- and D$4'$-branes are listed in Table \ref{table:BraneDirections}.  As we take the distance between the D4- and D$4'$-branes (along direction 9) to be large, the fundamental strings suspended between D4 and D$4'$ become non-dynamical. The boundaries of the fundamental strings on the D4-branes realize the line operators in the 5d gauge theory, whose positions are fixed by the D$4'$-brane. The Witten indices of the D0-D4-D$4'$ quantum mechanics were studied in \cite{Kim:2016qqs}. The D$4'$-brane introduces additional degrees of freedom coming from the D0-D$4'$ strings and D4-D$4'$ strings, which are listed in Table \ref{table:D0-D$4'$}.
\begin{table}\centering
\begin{tabular}{|c|c|c|c|}\hline
strings &${\cal N}=4$ multiplets & fields & {\footnotesize SU(2)$_-\times$SU(2)$_+\times$SU(2)$^R_-\times$SU(2)$^R_+$} 
\\ \hline
\multirow{3}{*}{D0-D$4'$ strings}& \multirow{2}{*}{twisted hyper} & scalar & $({\bf 1},{\bf 1},{\bf 1},{\bf 2})$
\\ \cline{3-4}
& & fermions & $({\bf 1},{\bf 2},{\bf 1},{\bf 1})$
\\ \cline{2-4}
& Fermi & fermions & $({\bf 2},{\bf 1},{\bf 1},{\bf 1})$
\\ \hline
D4-D$4'$ strings & Fermi & fermions & $({\bf 1},{\bf 1},{\bf 1},{\bf 1})$
\\ \hline
\end{tabular}
\caption{The quantum mechanics fields from the D0-D$4'$ strings and D4-D$4'$ strings.}
\label{table:D0-D$4'$}
\end{table}
The D0-D$4'$ and D4-D$4'$ strings are charged under the U(1) symmetry associated with the D$4'$-brane. We denote the fugacity of the U(1) by $x = e^{-M}$. Up to normalization, the distance between the D4- and D$4'$-branes is identified with $\log |x|$. It was shown in \cite{Kim:2016qqs} that the Witten indices admit a finite series expansion in the fugacity $x$, and the $k$-th order term in the expansion receives contributions from Wilson line in the $k$-th anti-symmetric representation of SU$(N)$.

Following \cite{Tong:2014cha, Kim:2016qqs}, we consider the D0-D4-D$4'$-D8/O8 quantum mechanics.\footnote{The D$2-$brane that we used in the construction of the ray operators in \tabref{table:BraneDirections} doesn't exist anymore when we radially quantize the theory on $S^4$ and focus on one of the two poles of $S^4$, to which the ray operator is mapped. Hence, we do not include D$2$ brane in the index calculation on the South Pole.} The Witten index of this quantum mechanics can be computed by a contour integral similarly to the way the index of the D$0$-D$4$-D$8$/O$8$ was calculated, as reviewed in \secref{sec:D0-D4-D8/O8}. The only modification required is to include the contributions from the D0-D$4'$ strings and D4-D$4'$ strings to the integrand of the $\phi$-contour integral \eqref{eqn:PhiIntegral}. It is easy to obtain the one-loop determinant of the D0-D$4'$ strings without any additional calculations. If we exchange the four-planes $\R^{1234}$ and $\R^{5678}$, the orientations of the D4- and D$4'$-branes are interchanged while the orientations of the D0-branes and D8/O8 singularity remain the same. One can also see from Table \ref{table:D0-D4-D8/O8} and Table \ref{table:D0-D$4'$} that when exchanging the SU(2)$_-\times$SU(2)$_+$ rotation symmetry of $\R^{1234}$ with the SU(2)$^R_-\times$SU(2)$^R_+$ rotation symmetry of $\R^{5678}$, the D0-D4 strings switch roles with the D0-D$4'$ strings while the other types of strings listed in the tables remain unchanged. To proceed, we reparametrize the chemical potentials associated to the spacetime rotations and introduce new chemical potentials $\epsilon_1, \dots,\epsilon_4$ by setting
\ie
\epsilon_+={\epsilon_1 + \epsilon_2\over 2},~~~\epsilon_-={\epsilon_1-\epsilon_2\over 2},
~~~m = {\epsilon_3 - \epsilon_4\over 2},
\fe
with the condition $\epsilon_1+\epsilon_2+\epsilon_3+\epsilon_4=0$. The $\epsilon_1$, $\epsilon_2$, $\epsilon_3$, $\epsilon_4$ are the chemical potentials associated to the rotations of the two-planes $\R^{12}$, $\R^{34}$, $\R^{56}$, $\R^{78}$. The exchange $\epsilon_1\leftrightarrow\epsilon_3$ combined with $\epsilon_2\leftrightarrow\epsilon_4$ corresponds to the exchange $\epsilon_+\leftrightarrow-\epsilon_+$ combined with $\epsilon_-\leftrightarrow m$. The one-loop determinant of the D0-D$4'$ strings is therefore obtained from the one-loop determinant of the D0-D4 strings \eqref{eqn:D0-D4Integrand}, by performing this simple substitution. The result is
\ie
&{\cal Z}^{+,\, k=2n+\chi}_{\text{\tiny D0-D$4'$}}=\bigg(\frac{2\sinh\frac{\pm M-\epsilon_-}{2}}{2\sinh\frac{\pm M-\epsilon_+}{2}}\bigg)^\chi\prod^n_{I=1}\frac{2\sinh\frac{\pm\phi_I\pm M-\epsilon_-}{2}}{2\sinh\frac{\pm\phi_I\pm M-\epsilon_+}{2}},
\\
&{\cal Z}^{-,\,k=2n+1}_{\text{\tiny D0-D$4'$}}=\frac{2\cosh\frac{\pm M-\epsilon_-}{2}}{2\cosh\frac{\pm M-\epsilon_+}{2}}\prod^{n}_{I=1}\frac{2\sinh\frac{\pm\phi_I\pm M-\epsilon_-}{2}}{2\sinh\frac{\pm\phi_I\pm M-\epsilon_+}{2}},
\\
&{\cal Z}^{-,\,k=2n}_{\text{\tiny D0-D$4'$}}=\frac{2\sinh(\pm M-\epsilon_-)}{2\sinh(\pm M-\epsilon_+)}\prod^{n-1}_{I=1}\frac{2\sinh\frac{\pm\phi_I\pm M-\epsilon_-}{2}}{2\sinh\frac{\pm\phi_I\pm M-\epsilon_+}{2}},
\fe
where we have also replaced the chemical potential $\alpha$ associated to the Sp(1) symmetry on the D4-brane with the chemical potential $M$ associated to the Sp(1) symmetry on the D$4'$-brane, and we used the shorthand notation of \cite{Hwang:2014uwa}, where $\sinh(\pm A\pm B\cdots)$ represents the product of sinh's of arguments with all possible sign combinations. (See \appref{app:integrand} for more details.)

The D4-D$4'$ string is a single fermion in the bifundamental representation of the Sp(1)$\times$Sp(1) symmetry, or equivalently in the vector representation of Spin(4) $\cong$ Sp(1)$\times$Sp(1).  The zero modes of the four components of the fermionic field in the vector representation form a 4d Clifford algebra, and the ground states form a spinor representation of the Clifford algebra. One then easily reads off the one-loop determinant of the D4-D$4'$ string,
\ie\label{eqn:D4-D$4'$}
{\cal Z}_{\text{\tiny D4-D$4'$}}= 2\cosh M - 2\cosh\alpha=2\sinh\tfrac{\pm\alpha-M}{2},
\fe
where the shorthand notation of \cite{Hwang:2014uwa}, reviewed in \appref{app:integrand}, was used again.

The integrands ${\cal Z}_{\text{\tiny D0-D$4'$}}$ and ${\cal Z}_{\text{\tiny D4-D$4'$}}$ have the $x$-expansions
\ie
{\cal Z}_{\text{\tiny D0-D$4'$}} = 1+ {\cal O}(x),~~~{\cal Z}_{\text{\tiny D4-D$4'$}} = {1\over x} - \chi_{\bf 2}(w) + x.
\fe
Plugging this into the $\phi$-contour integral and the integration formula \eqref{eqn:ROI}, one can see that the leading ${\cal O}(\frac{1}{x})$ order term in the expansion of the generating function ${\cal I}_{\text{\tiny all}}$ gives the superconformal index ${\cal I}_{\text{\tiny SCI}}$ and the ``naive'' expression \eqref{eqn:WRnaive} for the Wilson ray index ${\cal I}^{\text{\tiny Wilson ray}}_{\bf 2}$ contributes to ${\cal I}_{\text{\tiny ray}}$ in the ${\cal O}(1)$ order term of the expansion, as shown in \eqref{eqn:IallExpansion}. However, one should be cautious, because the $x$-expansion in general does not commute with the $\phi$-contour integral, and the above discussion should be just taken as heuristics. In the next section, we present the results for the ray operator indices obtained by evaluating the integrals \eqref{eqn:PhiIntegral} and \eqref{eqn:ROI} and expanding the generating function ${\cal I}_{\text{\tiny all}}$. We will demonstrate that the ray operator indices contain the ``naive'' Wilson ray operator indices and exhibit the $E_n$ symmetry.
%
\subsection{Ray operator indices}
\label{subsec:IndexRays}
We computed the ray operator indices up to ${\cal O}(t^5)$ order in the $t$-expansion, which receives contributions from up to instanton number five. For simplicity, except for the case $N_f=1$, we turn off all the SO($2N_f$) fugacities, and leave only the fugacity $q$ associated to the U(1) instanton number symmetry. We list our results for each value of $N_f$ below, including the correction $q^{-2/(8-N_f)}$ discussed in \secref{subsec:shiftI}.
Note that for $1\le N_f\le 6$ (i.e., $E_2,\dots,E_7$) the leading order term in the index is ${\cal O}(t)$, which corresponds to a doublet of the diagonal subgroup of SU(2)${}_{+}\times$SU(2)${}_R$, according to the discussion in \secref{subsec:StatesR}. Furthermore, for $2\le N_f\le 6$ the coefficient of the ${\cal O}(t)$ term is a character of a minuscule representation of $E_{N_f+1}$ (i.e., a representation whose weights form a single orbit of the Weyl group). Thus, these terms appear to capture the operators that generalize \eqref{eqn:WilsonSegment}, with a minuscule representation of $E_{N_f+1}$ playing the role of the fundamental representation of the flavor symmetry in an ordinary gauge theory coupled to quarks.
For convenience, we recall that
$$
t\equiv e^{-\epsilon_{+}},
\quad
u\equiv e^{-\epsilon_{-}},
$$
are the fugacities that couple to $J_{+}+J_R$ and $J_{-}$, respectively.

\subsubsection*{Ray operator index in $E_2=${\normalfont SU(2)}$\times${\normalfont U(1)} theory}
\ie
q^{-{2\over 7}}{\cal I}_{\text{\tiny ray}}&(t,u,m_\ell,q)=
\left[z^{4\over 7} + z^{-{3\over 7}}\chi_{\bf 2}(y)\right] t+z^{-{3\over 7}}\chi_{\bf 4}(y) t^3 
\\
&+\chi_{{\bf 2}}(u)\left[z^{-{3\over 7}}\left(\chi_{\bf 4}(y) + 2\chi_{\bf 2}(y)\right) + z^{4\over 7}\left(\chi_{\bf 3}(y)+ 1\right)\right]t^4
\\
&+\left[-z^{-{10\over 7}}\chi_{\bf 3}(y) + z^{-{3\over 7}}\left[\chi_{\bf 3}(u)(\chi_{\bf 4}(y) + 3\chi_{\bf 2}(y))+\chi_{\bf 6}(y)+\chi_{\bf 2}(y)\right]\right.
\\
&~~~+ z^{4\over 7}\left[\chi_{\bf 3}(u)(\chi_{\bf 3}(y) + 2)+1\right] - z^{11\over 7}\chi_{\bf 2}(y)\Big]t^5 + {\cal O}(t^6)
\fe
Here we defined the fugacities $y$ and $z$, which correspond to the SU(2) and U(1) factors of $E_2$ respectively. They are related to the fugacities $q$ and $y_1$ [where $y_1$ is associated with the flavor SO(2) symmetry of the $N_f=1$ 5d SYM] by \cite{Kim:2012gu},
\ie\label{eqn:yzqy1}
y^2 = q y_1,~~z^2 = {y^7_1\over q}.
\fe
(See \appref{app:E2} for more details.)
Note that the prefactor $q^{\frac{2}{7}}$ is $q^{2/(8-N_f)}$, in accordance with the shift in \eqref{eqn:qQshift}. The U(1) charge $\qQ$ of all the ray operators (as captured by $z$) is in $\tfrac{4}{7}+\Z$, and the SU$(2)$ spin $j$ [encoded in the character $\chi_{2j+1}(y)$] is integer (half-integer) when  
$\qQ-\tfrac{4}{7}$ is even (odd).

\subsubsection*{Ray operator index in $E_3=$ {\normalfont SU(3)}$\times${\normalfont SU(2)} theory}
\ie
q^{-{1\over 3}}{\cal I}_{\text{\tiny ray}}&(t,u,m_\ell,q)=\chi^{E_3}_{[1,0,1]} t+\left[\chi^{E_3}_{[2,1,1]}+\chi^{E_3}_{[1,0,3]}\right] t^3+\chi_{{\bf 2}}(u)\left[\chi^{E_3}_{[1,0,3]}+\chi^{E_3}_{[2,1,1]}+\chi^{E_3}_{[0,2,1]}\right]t^4
\\
&+\left[\chi_{{\bf 3}}(u)\big(\chi^{E_3}_{[1,0,1]}+\chi^{E_3}_{[1,0,3]}+\chi^{E_3}_{[2,1,1]}+\chi^{E_3}_{[0,2,1]}\big)+\chi^{E_3}_{[3,2,1]}+\chi^{E_3}_{[1,0,7]}\right]t^5+{\cal O}(t^6)
\fe
The relevant $E_3$ characters are as follows:
\ie
&\chi^{E_3}_{[1,0,1]}={4\over q^{1/3}}+2q^{2/3},
\\
&\chi^{E_3}_{[1,0,3]}={8\over q^{1/3}}+4q^{2/3},
\\
&\chi^{E_3}_{[1,0,7]}={16\over q^{1/3}}+8q^{2/3},
\\
&\chi^{E_3}_{[2,1,1]}={6\over q^{4/3}} + {12\over q^{1/3}} + 8 q^{2/3} + 4 q^{5/3},
\\
&\chi^{E_3}_{[0,2,1]}={2\over q^{4/3}}+{4\over q^{1/3}}+6q^{2/3},
\\
&\chi^{E_3}_{[3,2,1]}={8\over q^{7/3}}+{16\over q^{4/3}}+{24\over q^{1/3}}+18q^{2/3}+12q^{5/3}+6q^{8/3}.
\fe
Our notation for $E_3$ characters $\chi^{E_3}_{[a,b,c]}$ is equivalent to the product $\chi^{\rm SU(3)}_{[a,b]}\chi^{\rm SU(2)}_{c+1}$, where $j=c/2$ is the spin of the SU(2) representation, and ${[a,b]}$ denotes an SU(3) representation with Young diagram
\vskip 12pt
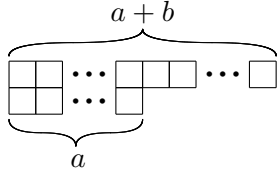
\begin{picture}(400,50)
\put(100,20){\begin{picture}(0,0)

\multiput(0,0)(0,10){2}{\begin{picture}(0,0)

\multiput(0,0)(10,0){2}{\begin{picture}(0,0)
\put(0,0){\line(0,1){10}}
\put(0,0){\line(1,0){10}}
\put(10,0){\line(0,1){10}}
\put(0,10){\line(1,0){10}}
\end{picture}}

\put(40,0){\begin{picture}(0,0)
\put(0,0){\line(0,1){10}}
\put(0,0){\line(1,0){10}}
\put(10,0){\line(0,1){10}}
\put(0,10){\line(1,0){10}}
\end{picture}}

\multiput(25,5)(5,0){3}{\circle*{2}}

\end{picture}}

\multiput(50,10)(10,0){2}{\begin{picture}(0,0)
\put(0,0){\line(0,1){10}}
\put(0,0){\line(1,0){10}}
\put(10,0){\line(0,1){10}}
\put(0,10){\line(1,0){10}}
\end{picture}}

\multiput(75,15)(5,0){3}{\circle*{2}}

\put(90,10){\begin{picture}(0,0)
\put(0,0){\line(0,1){10}}
\put(0,0){\line(1,0){10}}
\put(10,0){\line(0,1){10}}
\put(0,10){\line(1,0){10}}
\end{picture}}

\qbezier(0,-2)(0,-7)(12,-7)
\qbezier(12,-7)(25,-7)(25,-12)
\qbezier(25,-12)(25,-7)(38,-7)
\qbezier(38,-7)(50,-7)(50,-2)
\put(23,-20){$a$}

\qbezier(0,22)(0,27)(24,27)
\qbezier(24,27)(50,27)(50,32)
\qbezier(50,32)(50,27)(76,27)
\qbezier(76,27)(100,27)(100,22)
\put(38,35){$a+b$}
\end{picture}}
\end{picture}
\vskip 12pt
Note that the spin $j=c/2$ is always half integral, and the number of boxes in the Young diagram is always $2\pmod{3}$. This corresponds to charge $5\pmod{6}$ under the $\Z_6\cong\Z_3\times\Z_2$ center of $E_3$.
Note also that the coefficient of the ${\cal O}(t)$ term is the character of the minuscule representation $(\rep{3},\rep{2})$ of $E_3\cong$ SU(3)$\times$SU(2).

\subsubsection*{Ray operator index in $E_4=$ {\normalfont SU(5)} theory}
\ie
q^{-{2\over 5}}{\cal I}_{\text{\tiny ray}}&(t,u,m_\ell,q)=
\chi^{E_4}_{[0,1,0,0]} t+\chi^{E_4}_{[1,1,0,1]} t^3
\\
&+\chi_{{\bf 2}}(u)\left[\chi^{E_4}_{[1,1,0,1]}+\chi^{E_4}_{[2,0,0,0]}+\chi^{E_4}_{[0,0,1,1]}+\chi^{E_4}_{[0,1,0,0]}\right]t^4
\\
&+\left[\chi_{{\bf 3}}(u)\big(\chi^{E_4}_{[1,1,0,1]}+\chi^{E_4}_{[2,0,0,0]}+\chi^{E_4}_{[0,0,1,1]}+2\chi^{E_4}_{[0,1,0,0]}\big)\right.
\\
&~~~\left.+\chi^{E_4}_{[2,1,0,2]}+\chi^{E_4}_{[1,1,0,1]}+\chi^{E_4}_{[2,0,0,0]}+\chi^{E_4}_{[0,0,1,1]}+3\chi^{E_4}_{[0,1,0,0]}\right]t^5+{\cal O}(t^6)
\fe
The relevant $E_4$ characters are as follows:
\ie
&\chi^{E_4}_{[0,1,0,0]}={6\over q^{2/5}}+4q^{3/5},
\\
&\chi^{E_4}_{[0,0,1,1]}={4\over q^{7/5}} + {16\over q^{2/5}} + 20 q^{3/5},
\\
&\chi^{E_4}_{[1,1,0,1]}={20\over q^{7/5}} + {80\over q^{2/5}} + 60 q^{3/5} + 15 q^{8/5},
\\
&\chi^{E_4}_{[2,0,0,0]}={10\over q^{2/5}} + 4 q^{3/5} + q^{8/5},
\\
&\chi^{E_4}_{[2,1,0,2]}={45\over q^{12/5}} + {180\over q^{7/5}} + {450\over q^{2/5}} + 360 q^{3/5} + 144 q^{8/5} + 36 q^{13/5}.
\fe
The representation $[a,b,c,d]$ corresponds to a Young diagram with rows of lengths $a+b+c+d, a+b+c, a+b, a$. Note that the representations have Young diagrams with total number of boxes $4a+3b+2c+d=3,8,13,\ldots$. Thus, under the $\Z_5$ center they charge $3\pmod{5}$, as promised in \secref{subsec:shiftI}.
Note also that the coefficient of the ${\cal O}(t)$ term is the character of the minuscule representation $\overline{\rep{10}}$ of $E_4\cong$ SU(5).

\subsubsection*{Ray operator index in $E_5=$ {\normalfont SO(10)} theory}
\ie
q^{-{1\over 2}}{\cal I}_{\text{\tiny ray}}&(t,u,m_\ell,q)=\chi^{E_5}_{[0,0,0,0,1]} t+\chi^{E_5}_{[0,1,0,0,1]} t^3+\chi_{{\bf 2}}(u)\left[\chi^{E_5}_{[1,0,0,1,0]}+\chi^{E_5}_{[0,1,0,0,1]}+\chi^{E_5}_{[0,0,0,0,1]}\right]t^4
\\
& +  \left[\chi_{{\bf 3}}(u)\big(\chi^{E_5}_{[1,0,0,1,0]}+\chi^{E_5}_{[0,1,0,0,1]}+2\chi^{E_5}_{[0,0,0,0,1]}\big)+\chi^{E_5}_{[0,2,0,1,0]}+\chi^{E_5}_{[0,0,0,0,1]}\right]t^5 + {\cal O}(t^6)
\fe
The relevant $E_5$ characters are as follows:
\ie
&\chi^{E_5}_{[0,0,0,0,1]}={8\over q^{1/2}}+8q^{1/2},
\\
&\chi^{E_5}_{[0,1,0,0,1]}={56\over q^{3/2}} + {224\over q^{1/2}}+224q^{1/2} + 56 q^{3/2},
\\
&\chi^{E_5}_{[1,0,0,1,0]}={8\over q^{3/2}}+{64\over q^{1/2}}+64q^{1/2} + 8q^{3/2},
\\
&\chi^{E_5}_{[0,2,0,1,0]}={224\over q^{5/2}} + {1120\over q^{3/2}} + {2688\over q^{1/2}} + 2688q^{1/2}  + 1120q^{3/2} + 224 q^{5/2}.
\fe
Recall that the root lattice is a sublattice of index $4$ in the weight lattice of $E_5$. The quotient of the weight lattice by root lattice can be identified with the Pontryagin dual of the $\Z_4\subset E_5$ center, and all the weights appearing in the characters above project to the same generator of $\Z_4$. In other words, there is a natural assignment of an additive $\Z_4$ charge to every weight, with roots having charge $0$, and it is not hard to check that all the weights appearing above have the same nonzero $\Z_4$ charge, which is $\pm 1$ (depending on convention). As discussed in \secref{subsec:shiftI}, referring to the ``SO(8)$\times$U(1)'' subgroup, the value of the $\Z_4$ charge when taken mod $2$ corresponds to the U(1) charge mod $2$. The fact that all $q$ powers in the ray operator index are half integers means that the U(1) charge is odd, and this confirms that the $\Z_4$ charge is $\pm1\pmod{4}$.
Note also that the coefficient of the ${\cal O}(t)$ term is the character of the minuscule representation $\rep{16}$ of $E_5\cong$ SO(10).

\subsubsection*{Ray operator index in $E_6$ theory}
\ie
q^{-{2\over 3}}{\cal I}_{\text{\tiny ray}}&(t,u,m_\ell,q)=\chi^{E_6}_{[0,0,0,0,0,1]} t+\chi^{E_6}_{[0,1,0,0,0,1]} t^3+\chi_{{\bf 2}}(u)\left[\chi^{E_6}_{[0,0,0,0,0,1]}+\chi^{E_6}_{[0,1,0,0,0,1]}+\chi^{E_6}_{[0,0,1,0,0,0]}\right]t^4
\\
&+\left[\chi_{{\bf 3}}(u)\big(2\chi^{E_6}_{[0,0,0,0,0,1]}+\chi^{E_6}_{[0,1,0,0,0,1]}+\chi^{E_6}_{[0,0,1,0,0,0]}\big)+\chi^{E_6}_{[0,2,0,0,0,1]}+\chi^{E_6}_{[0,0,0,0,0,1]}\right]t^5+{\cal O}(t^6)
\fe
The relevant $E_6$ characters are as follows:
\ie
&\chi^{E_6}_{[0,0,0,0,0,1]}={10\over q^{2/ 3}}+16q^{1/ 3}+q^{4/ 3},
\\
&\chi^{E_6}_{[0,0,1,0,0,0]}={16\over q^{5/ 3}} + {130\over q^{2/ 3}} + 160 q^{1/ 3} + 45 q^{4/ 3},
\\
&\chi^{E_6}_{[0,1,0,0,0,1]}={144\over q^{5/ 3}}+{576\over q^{2/ 3}}+736 q^{1/ 3} + 256 q^{4/ 3}+ 16 q^{7/ 3},
\\
&\chi^{E_6}_{[0,2,0,0,0,1]}={1050\over q^{8/ 3}}+{5712\over q^{5/ 3}}+{13506\over q^{2/ 3}}+ 15696 q^{1/ 3} + 8226 q^{4/ 3}+ 2016 q^{7/ 3} + 126 q^{10/ 3}.
\fe
The root lattice of $E_6$ is a sublattice of index $3$ in the weight lattice. The quotient of the weight lattice by root lattice can be identified with the Pontryagin dual of the $\Z_3\subset E_6$ center, and again all the weights appearing in the characters above project to the same generator of $\Z_3$. This is consistent with the discussion in \secref{subsec:shiftI}, and indeed, as promised there, all the $E_6$ characters that appear in the index of ray operators decompose under SO(10)$\times$U(1) in such a way that the powers of $q$ (which are proportional to the U(1) charge) take values in $\tfrac{1}{3}+\Z$.
Note also that the coefficient of the ${\cal O}(t)$ term is the character of the minuscule representation $\rep{27}$ of $E_6$.

\subsubsection*{Ray operator index in $E_7$ theory}
\ie
q^{-1}{\cal I}_{\text{\tiny ray}}&(t,u,m_\ell,q)=
\chi^{E_7}_{[0,0,0,0,0,0,1]} t+\chi^{E_7}_{[1,0,0,0,0,0,1]}t^3
\\
&+\chi_{{\bf 2}}(u)\left[\chi^{E_7}_{[0,0,0,0,0,0,1]}+\chi^{E_7}_{[0,1,0,0,0,0,0]}+\chi^{E_7}_{[1,0,0,0,0,0,1]}\right]t^4
\\
& + \left[\chi_{{\bf 3}}(u)\big(2\chi^{E_7}_{[0,0,0,0,0,0,1]}+\chi^{E_7}_{[0,1,0,0,0,0,0]}+\chi^{E_7}_{[1,0,0,0,0,0,1]}\big)\right.
\\
&~~~\left.+\chi^{E_7}_{[2,0,0,0,0,0,1]}+\chi^{E_7}_{[0,0,0,0,0,0,1]}\right]t^5 +{\cal O}(t^6)
\fe
The relevant $E_7$ characters are listed as follows
\ie
&\chi^{E_7}_{[0,0,0,0,0,0,1]} = {12\over q}+32 + 12 q,
\\
&\chi^{E_7}_{[0,1,0,0,0,0,0]}={32\over q^2} + {232\over q} + 384 + 232 q + 32 q^2,
\\
&\chi^{E_7}_{[1,0,0,0,0,0,1]}={12\over q^3} + {384\over q^2} + {1596\over q} + 2496 + 1596 q+384 q^2+12 q^3,
\\
&\chi^{E_7}_{[2,0,0,0,0,0,1]}={12\over q^5} + {384\over q^4} + {6348\over q^3} + {31008\over q^2} + {73536\over q} + 97536
\\
&~~~~~~~~~~~~~~~~~~+ 73536 q+31008 q^2+ 6348 q^3 + 384q^4 + 12q^5.
\fe
It is not hard to check that all the weights that appear in the characters above do not belong to the root lattice (i.e., they are not representations of the adjoint form of $E_7$). Since the  root lattice is a sublattice of index $2$ in the weight lattice of $E_7$, we see that all the $E_7$ representations of ray operators are odd under the $\Z_2$ center.
Note also that the coefficient of the ${\cal O}(t)$ term is the character of the minuscule representation $\rep{56}$ of $E_7$.
\subsubsection*{Ray operator index in $E_8$ theory}

The case $N_f=7$ poses a special challenge, because we do not have a consistent result for the South Pole contribution $\ZinstLine$ to the partition function \eqref{eqn:ROI}. The problem, as we discussed below equation \eqref{eqn:LineZoverZ}, is that a direct computation of $\ZinstLine$, following the ideas developed in \cite{Hwang:2014uwa}, yields a result that depends on the SU(2)$^R_{-}$ fugacity $v$.
Nevertheless, it is instructive to look at the result of the integral formula \eqref{eqn:ROI}, after substituting for $\ZinstLine$ the problematic formula \eqref{eqn:LineZoverZ}.
With $\chi_{\rep{2}}(v)\equiv v + \frac{1}{v}$, 
and ${\cal Z}_{\text{\tiny inst}}$ denoting the instanton partition function \eqref{InstantonFromQM} without the line, we find
\ie\label{eqn:RayIndexE8}
q^{-2}{\cal I}^{\text{\tiny(calculated)}}_{\text{\tiny ray}}(t,u,v,m_\ell,q)
=\chi_{\rep{2}}(v){\cal I}_{\text{\tiny SCI}}(t,u,m_\ell,q)
+{\cal I}_{\text{\tiny $v$-independent}}(t,u,m_\ell,q),
\fe
where ${\cal I}_{\text{\tiny SCI}}$ is the index of local operators given in \eqref{SCI}, and 
\ie\label{eqn:RayIndexE8nov}
{\cal I}_{\text{\tiny $v$-independent}} &=
\big(1+\chi^{E_8}_{[0,0,0,0,0,0,0,1]}\big) t+\chi_{\rep{2}}(u)t^2+\left(\chi^{E_8}_{[0,0,0,0,0,0,0,1]}+\chi^{E_8}_{[0,0,0,0,0,0,1,0]}+\chi^{E_8}_{[0,0,0,0,0,0,0,2]}\right)t^3\
\\
&+\left\{\chi_{{\bf 2}}(u)+\chi_{{\bf 2}}(u)\big(3\chi^{E_8}_{[0,0,0,0,0,0,0,1]}+\chi^{E_8}_{[1,0,0,0,0,0,0,0]}+\chi^{E_8}_{[0,0,0,0,0,0,0,2]}+\chi^{E_8}_{[0,0,0,0,0,0,1,0]}\big)\right\}t^4
\\
&+\left\{2+2\chi^{E_8}_{[0,0,0,0,0,0,0,1]}+\chi^{E_8}_{[0,0,0,0,0,0,1,1]}+\chi^{E_8}_{[0,0,0,0,0,0,0,2]}+\chi^{E_8}_{[0,0,0,0,0,0,0,3]}\right.
\\
&~~~\left.+\chi_{{\bf 3}}(u)\big(2+4\chi^{E_8}_{[0,0,0,0,0,0,0,1]}+\chi^{E_8}_{[1,0,0,0,0,0,0,0]}+\chi^{E_8}_{[0,0,0,0,0,0,0,2]}+\chi^{E_8}_{[0,0,0,0,0,0,1,0]}\big)\right\}t^5+{\cal O}(t^6).
\fe
The relevant $E_8$ characters are listed as follows
\ie
&\chi_{[0,0,0,0,0,0,0,1]}={14\over q^2}+{64\over q}+92+ 64q +14q^2,
\\
&\chi_{[1,0,0,0,0,0,0,0]}={1\over q^4}+{64\over q^3}+{378\over q^2}+{896\over q}+1197+ 896q+ 378 q^2+ 64 q^3+q^4,
\\
&\chi_{[0,0,0,0,0,0,1,0]}={91\over q^4}+{896\over q^3}+{3290\over q^2}+{6720\over q}+8386+ 6720q+ 3290 q^2+ 896 q^3+91 q^4,
\\
&\chi_{[0,0,0,0,0,0,0,2]}={104\over q^4}+{832\over q^3}+{2990\over q^2}+{5888\over q}+7372+ 5888q+ 2990 q^2+ 832 q^3+104 q^4,
\\
&\chi_{[0,0,0,0,0,0,1,1]}={896\over q^6}+{11584\over q^5}+{65792\over q^4}+{221248\over q^3}+{496768\over q^2}+{791168\over q}+921088
\\
&~~~~~~~~~~~~~~~~~~~~~+791168q + 496768 q^2 + 221248 q^3 + 65792 q^4 + 11584 q^5 + 896 q^6,
\\
&\chi_{[0,0,0,0,0,0,0,3]}={546\over q^6}+{5824\over q^5}+{30394\over q^4}+{98176\over q^3}+{214474\over q^2}+{336960\over q}+390377
\\
&~~~~~~~~~~~~~~~~~~~~~+336960 q + 214474 q^2 + 98176 q^3 + 30394 q^4 + 5824 q^5 + 546 q^6.
\fe
We computed the contribution to the formula \eqref{eqn:RayIndexE8nov} up to instanton number five. The ${\cal O}(t^5)$ order of \eqref{eqn:RayIndexE8nov} receives contribution from higher instanton number, and we completed the formula \eqref{eqn:RayIndexE8nov} ``by hand'' using the property $\chi^{E_8}_{\bf R}(q) = \chi^{E_8}_{\bf R}(q^{-1})$ of the $E_8$ characters.

Note that since ${\cal I}_{\text{\tiny SCI}}=1+{\cal O}(t^2)$, the expression \eqref{eqn:RayIndexE8} starts with a $t$-independent term $\chi_{\rep{2}}(v)$. But a ray operator index is forbidden from having such a term, according to the discussion at the end of \secref{subsec:StatesR}, since it would require the existence of a BPS ray operator with $J_{+}+J_R=0$. 
More generally, the first term on the RHS of \eqref{eqn:RayIndexE8} suggests that there are unwanted SU(2)${}_{-}^R$ doublet states that contribute to the partition function \eqref{eqn:LineZoverZ}.
It is tempting to drop the $\chi_{\rep{2}}(v){\cal I}_{\text{\tiny SCI}}$ term entirely from the ray index and keep only the $v$-independent term, but we have not found a satisfactory argument for this ad-hoc prescription, and it is not clear whether any additional SU(2)${}_{-}^R$ singlets should be dropped as well, or not.

$N_f=7$ is special, because the parameter $\mFlux=8-N_f$ is exactly $1$ in this case, which makes it possible for the F1 that appeared in the construction of the line and ray operators in \secref{sec:LRO} to end on a D$0$-brane instead of the D$4$-brane. The F1 can thus be ``screened'', and the $x^9$ coordinate of the D$0$-brane is a free parameter, which gives rise to a continuum, in the absence of the D$4$'-brane.\footnote{The mass of the D$0$-brane increases linearly with $x^9$, but this effect is canceled by the decreasing length of F$1$, and so there is no potential.}
We do not understand why this effect  creates the $v$-dependent terms, but we suspect that it is part of the problem.

\section{Discussion}
\label{sec:Disc}

We have extended the analysis of \cite{Hwang:2014uwa} by calculating the index of ray operators in 5d $E_n$ SCFTs for $n=2,\dots,7$. We converted the problem to a partition function of the SCFT on S$^4\times$S$^1$ with a Wilson loop along S$^1$ and with twisted boundary conditions parameterized by the various fugacities. Following \cite{Hwang:2014uwa}, we provided evidence that the manifest SO$(2n-2)$ flavor symmetry combines with the U(1) symmetry associated with the conserved instanton charge to form a subgroup of an enhanced $E_n$ ``flavor'' symmetry, as predicted in \cite{Seiberg:1996bd}.
Our index reveals $E_n$ representations that do not appear in the superconformal index of local operators. These are representations with weights that are not in the root lattice of $E_n$, and the ray operators are charged under the nontrivial center of $E_n$. 
For $n=8$ we encountered a problem with the calculation of the contribution of zero-size instantons to the ray index. The prescription that we followed for calculating the Nekrasov partition function in the presence of a Wilson loop
does not appear to yield a result that factorizes properly into field theory modes and modes that are decoupled from the D$4$-brane.
We do not know the reason for this inconsistency, but we suspect it has to do with the possibility for a fundamental string (that induces the line operator) to end on a D$0$-brane.

As in the work of \cite{Hwang:2014uwa}, a key ingredient in the calculation is the contribution of coincident zero-size instantons. In our case, the instantons are also coincident with the defect introduced by a Wilson loop, and we needed to regularize their contribution carefully. As we saw in \secref{subsec:IndexROwr}, merely localizing the Wilson loop on BPS configurations is not the right answer. Instead, we followed \cite{Tong:2014cha,Nekrasov:2015wsu,Kim:2016qqs} and rather than introducing the Wilson loop directly to the S$^4\times$S$^1$ partition function, we modified the Nekrasov partition function that captures the contribution of the zero-size instantons near the Wilson loop. The modified Nekrasov partition function is an index of the quantum mechanics of D$0$-branes that probe a D$4$-D$8$/O$8$ system, and the Wilson loop was captured by introducing an additional D$4$-brane (denoted by D$4$') to the system so that after integrating out the (heavy) fermionic D$4$-D$4$' string modes, the Wilson loop is recovered.
That the final result (after inserting this modified Nekrasov partition function into the 5d index formula) reveals the expected hidden $E_n$ global symmetry lends credence to this resolution of zero-size instanton singularities in our context as well.
The modified Nekrasov partition function also appeared as part of Nekrasov's larger work \cite{Nekrasov:2015wsu} on {\it the qq-character.}

A better understanding of how the exceptional symmetry of the $E_n$ SCFTs arises is important both in its own right and since the $E_n$ SCFTs describe the low-energy degrees of freedom of M-theory near degenerations of Calabi-Yau manifolds \cite{Morrison:1996xf} and can also provide clues about the 6d $(1,0)$ SCFT with $E_8$ global symmetry. The 5d ray operators presumably descend from BPS cylinder operators in 6d, that is, surface operators associated with open surfaces with S$^1\times$$\R_{+}$ geometry. The AdS/CFT dual of such an operator, as well as the analysis of \secref{sec:LRO} suggest that the 1d boundary of these operators are ``labeled'' by a state of an $E_8$ affine Lie algebra at level $1$. This is the extended symmetry of the low-energy 2d CFT that described the M$2$-M$9$ intersection \cite{Horava:1995qa}. It would be interesting to examine further the relationship between 5d ray operators and 6d surface operators.

On the Coulomb branch of the $E_n$ theories the low-energy description is given by a single U(1) vector multiplet, and the ray operators that we counted in this work can act on the vacuum and create BPS states with one unit of charge. It would be interesting to explore the relation between the BPS spectrum of the $E_n$ theories on their Coulomb branch (computed in \cite{Klemm:1996hh,Minahan:1997ch,Kol:1998cf}) and the index that we calculated in this paper \cite{toAppear}. Indeed, for $n=2,3,4,5,6,7$, the net numbers of $\tfrac{1}{8}$BPS operators with $J_{+}+J_R=\tfrac{1}{2}$ and $J_{-}=0$ are $3,6,10,16,27,56$. These are precisely the numbers of isolated holomorphic curves of genus $0$ embedded in the del Pezzo surface $B_n$ \cite{Katz:1999xq}, and are a special case of the Gopakumar-Vafa invariants of the Calabi-Yau manifolds that enter the M-theory construction of the $E_n$ theories \cite{Gopakumar:1998ii,Gopakumar:1998jq}. It would be interesting \cite{toAppear} to explore the connection between ray operator indices generated by probing the D$4$-brane with more than one fundamental string and Gopakumar-Vafa invariants of higher genera and degrees, as computed in \cite{Katz:1999xq}.

Our results are also related to the elliptic genus of an E-string near a surface operator of the 6d $(1,0)$ $E_8$-theory. The E-string is the BPS string-like excitation of the 6d theory on the Coulomb branch. A single E-string is described by a left-moving $E_8$ chiral current algebra together with four noninteracting 2d bosons and right-moving fermions, but $k$ coincident E-strings have nontrivial 2d CFT descriptions with $(4,0)$ supersymmetry \cite{Haghighat:2014pva}.
It was shown in \cite{Kim:2014dza} that the intermediate steps in the computation of a 5d index for the $N_f=8$ case can be used to also compute the elliptic genus of $k$ E-strings (see also \cite{Haghighat:2014pva,Hohenegger:2015cba}).
More precisely, the index of the 5d theory on S$^4\times$S$^1$ is a contour integral over a complex variable $w$ that can be identified with the holonomy of a U(1) $\subset$ SU(2) gauge field on S$^1$. The integrand is a product of terms, one of which is a Nekrasov partition function whose $w^k$ coefficient yields the elliptic genus of $k$ E-strings.
Our computation of the index of ray operators also has a Nekrasov partition function ingredient, from which a modified E-string elliptic genus can be read off.
It counts bound states of $k$ E-strings and a 1+1d defect, introduced into the 6d theory via a BPS surface operator, and compactified on S$^1$.

Our calculation, which builds on the techniques developed in \cite{Kim:2012gu,Hwang:2014uwa}, uses an ordinary super Yang-Mills theory to capture properties of a strongly interacting SCFT. It joins a growing body of work that demonstrates that the manifestly nonrenormalizable Yang-Mills theories in dimensions $d>4$ still prove to be very useful in the right context. For example, \cite{Douglas:2010iu} proposed that 5d super Yang-Mills theory can describe the 6d $(2,0)$-theory, in \cite{Kim:2011mv} it was shown how Yang-Mills theory can be used to calculate a superconformal index for the 6d $(2,0)$-theory and reproduce its anomaly coefficient,
and in \cite{Chang:2014jta,Lin:2015zea} it was demonstrated that a 6d Yang-Mills theory can be used to calculate Little String Theory amplitudes.

The localization computation of the superconformal indices in \cite{Kim:2012gu,Hwang:2014uwa} requires deforming the 5d SYM in a way that keeps the indices invariant. In \secref{subsec:SCIfromSYM}, we demonstrated that the perturbative part of the indices \eqref{PertIndex} can be reproduced by directly counting the local gauge invariant operators in the 5d SYM.  One expects that the instanton contribution to the indices can be reproduced in a similar way, involving quantizing the moduli space of the instantons on S$^4$ and counting the instanton operators \cite{Lambert:2014jna,Rodriguez-Gomez:2015xwa,Tachikawa:2015mha,Bergman:2016avc} in 5d SYM. Similar problems have been studied in 3d Chern-Simons matter theories \cite{Kim:2009ia,Berenstein:2009sa,Kim:2010ac,Aharony:2015pla}, where partial success was achieved, and the superconformal indices were computed in certain monopole sectors by directly counting monopole operators.

\section*{Acknowledgements}
We are grateful to Oren Bergman, Joonho Kim, Ying-Hsuan Lin, Daniel Parker, Christian Schmid, Shu-Heng Shao, Cumrun Vafa, Yifan Wang, Ziqi Yan, and Xi Yin, for helpful advice, conversations and discussions.
This research was supported in part by the Berkeley Center of Theoretical Physics. OJG also wishes to thank the Aspen Center for Physics, which is supported by National Science Foundation grant PHY-1066293, where part of this work was completed.
The research of JO was supported by Kwanjeong Educational Foundation.

\begin{appendix}
\section{One-loop determinants in the D0-D4-D8/O8 quantum mechanics}
\label{app:integrand}

The one-loop determinants of the D0-D4-D8/O8 quantum mechanics fields, listed in \tabref{table:D0-D4-D8/O8}, were computed in \cite{Kim:2012gu,Hwang:2014uwa}.  The exact forms can be found in equations (3.42)-(3.50) of \cite{Hwang:2014uwa}, and we summarize them in this appendix, using the conventions of \cite{Hwang:2014uwa} whereby, for example, a term of the form $2\sinh(\pm A\pm B\pm C + D)$ should be interpreted as a product over eight terms (all combinations of $\pm$ signs):
\bear
\lefteqn{
2\sinh(\pm A\pm B\pm C + D)\rightarrow
256
\sinh(A+B+C+D)
\sinh(A+B-C+D)
}\nn\\ &&\qquad\qquad\qquad
\sinh(A-B+C+D)
\sinh(A-B-C+D)
\sinh(-A+B+C+D)
\nn\\ &&\qquad\qquad\qquad
\sinh(-A+B-C+D)
\sinh(-A-B+C+D)
\sinh(-A-B-C+D).
\nn
\eear

The one-loop determinants of the D0-D0 strings are given by
\ie\label{eqn:Zplus}
\hspace*{-1cm}{\cal Z}_{\text{\tiny D0-D0}}^{+,\,k=2n+\chi}=& \bigg[\Big(\prod_{I=1}^{n} 2\sinh{\tfrac{\pm \phi_I}{2}}\Big)^{\chi }\prod_{I<J}^{n} 2\sinh{ \tfrac{ \pm \phi_I \pm \phi_J}{2}} \bigg](2 \sinh{\epsilon_+})^n\Big(\prod_{I=1}^{n}2\sinh{ \tfrac{\pm \phi_I + 2\epsilon_+}{2}}\Big)^{\chi}\prod_{I < J}^{n}2\sinh{ \tfrac{\pm \phi_{I} \pm \phi_{J} + 2\epsilon_+}{2} }
\\
&\hspace*{-.5cm}\times\big(2 \sinh{\tfrac{\pm m - \epsilon_-}{2}}\big)^n \Big(\prod_{I=1}^{n}2\sinh{\tfrac{\pm\phi_I \pm m - \epsilon_-}{2}}\Big)^\chi \prod_{I<J}^{n}2\sinh{\tfrac{\pm\phi_I \pm \phi_J \pm m - \epsilon_-}{2}} 
\\
&\hspace*{-.5cm}\times {1\over \big(2\sinh{\frac{\pm m - \epsilon_+}{2}}\big)^{n+\chi}}\Big(\prod_{I=1}^{n} \frac{1}{2\sinh{\frac{\pm\phi_I \pm m - \epsilon_+}{2}}} \Big)^{\chi}\prod_{I=1}^{n} \frac{1}{2 \sinh{\frac{\pm 2\phi_I \pm m - \epsilon_+}{2}} } \prod_{I<J}^{n} \frac{1}{2\sinh{\frac{\pm\phi_I \pm \phi_J \pm m - \epsilon_+}{2}}}
\\
&\hspace*{-.5cm}\times{1\over \big(2 \sinh{\frac{\pm \epsilon_- + \epsilon_+ }{2}} \big)^{n+\chi}} \Big( \prod_{I=1}^{n} \frac{ 1} {2\sinh{ \frac{\pm \phi_I \pm \epsilon_- + \epsilon_+}{2}}}\Big)^{\chi}
\prod_{I=1}^{n} \frac{1 }{  2\sinh{ \frac{\pm 2\phi_{I} \pm \epsilon_- + \epsilon_+}{2}}  } \prod_{I < J}^{n} \frac{1 }{2\sinh{ \frac{\pm \phi_{I} \pm \phi_{J} \pm \epsilon_- + \epsilon_+}{2}}},
\fe
and
\ie\label{eqn:Zminusodd}
\hspace*{-1cm}{\cal Z}_{\text{\tiny D0-D0}}^{-,\,k=2n+1}=&\Big( \prod_I^{n} 2\cosh{\tfrac{\pm \phi_I}{2}} \prod_{I<J}^{n} 2\sinh{ \tfrac{ \pm \phi_I \pm \phi_J}{2}}\Big) (2 \sinh{\epsilon_+})^n\prod_{I=1}^{n} 2\cosh{ \tfrac{\pm \phi_I + 2\epsilon_+}{2}}\prod_{I < J}^{n}2\sinh{ \tfrac{\pm \phi_{I} \pm \phi_{J} + 2\epsilon_+}{2} }
\\
&\hspace*{-0.5cm}\times\big(2 \sinh{\tfrac{\pm m - \epsilon_-}{2}}\big)^n \prod_{I=1}^{n}2\cosh{\tfrac{\pm\phi_I \pm m - \epsilon_-}{2}} \prod_{I < J}^{n}2\sinh{\tfrac{\pm\phi_I \pm \phi_J \pm m - \epsilon_-}{2}}
\\
&\hspace*{-0.5cm}\times\frac{1}{\big(2\sinh{\frac{\pm m - \epsilon_+}{2}}\big)^{n+1}}  \prod_{I=1}^{n} \frac{1}{2\cosh{\frac{\pm\phi_I \pm m - \epsilon_+}{2}}2 \sinh{\frac{\pm 2\phi_I \pm m - \epsilon_+}{2}} }  \prod_{I<J}^{n} \frac{1}{2\sinh{\frac{\pm\phi_I \pm \phi_J \pm m - \epsilon_+}{2}}}
\\
&\hspace*{-0.5cm}\times \frac{1}{\big(2\sinh{ \frac{\pm \epsilon_- + \epsilon_+}{2}} \big)^{n+1} }  \prod_{I=1}^{n} \frac{ 1} {2\cosh{ \frac{\pm \phi_I \pm \epsilon_- + \epsilon_+  }{2}}  2\sinh{ \frac{\pm 2\phi_{I} \pm \epsilon_- + \epsilon_+}{2}}  } \prod_{I < J}^{n} \frac{ 1}{2\sinh{ \frac{\pm \phi_{I} \pm \phi_{J} \pm \epsilon_- + \epsilon_+}{2}}},
\fe
and
\ie\label{eqn:Zminuseven}
\hspace*{-2cm}{\cal Z}_{\text{\tiny D0-D0}}^{-,\,k=2n}=&\Big(\prod_{I<J}^{n-1} 2\sinh{ \tfrac{ \pm \phi_I \pm \phi_J}{2}}  \prod_I^{n-1} 2\sinh{(\pm \phi_I)} \Big) 2\cosh{\epsilon_+}(2 \sinh{\epsilon_+} )^{n-1}\prod_{I=1}^{n-1} 2\sinh{(\pm \phi_I + 2\epsilon_+) }  \prod_{I < J}^{n-1}  2\sinh{ \tfrac{\pm \phi_{I} \pm \phi_{J} + 2\epsilon_+}{2} }
\\
&\hspace*{-0.5cm}\times 2 \cosh{\tfrac{\pm m - \epsilon_-}{2}} \big(2 \sinh{\tfrac{\pm m - \epsilon_-}{2}}\big)^{n-1}\prod_{I=1}^{n-1}2\sinh{(\pm\phi_I \pm m - \epsilon_-)} \prod_{I<J}^{n-1}2\sinh{\tfrac{\pm\phi_I \pm \phi_J \pm m - \epsilon_-}{2}}
\\
&\hspace*{-0.5cm}\times \frac{1}{\big(2\sinh{\frac{\pm m - \epsilon_+}{2}}\big)^n 2\sinh{(\pm m - \epsilon_+)}} \prod_{I=1}^{n-1} \frac{1}{2\sinh{(\pm\phi_I \pm m - \epsilon_+)}\sinh{\frac{\pm 2\phi_I \pm m - \epsilon_+}{2}} } 
\prod_{I<J}^{n-1} \frac{1}{2\sinh{\frac{\pm\phi_I \pm \phi_J \pm m - \epsilon_+}{2}}}
\\
&\hspace*{-0.5cm}\times \frac{1}{\big(2\sinh{ \frac{\pm \epsilon_- + \epsilon_+}{2}} \big)^n 2\sinh{ (\pm \epsilon_- + \epsilon_+)}}  \prod_{I=1}^{n-1} \frac{1 } {2\sinh{ (\pm \phi_I \pm \epsilon_- + \epsilon_+)} 2\sinh{ \frac{\pm 2\phi_{I} \pm \epsilon_- + \epsilon_+}{2}}}\prod_{I < J}^{n-1} \frac{1}{2\sinh{ \frac{\pm \phi_{I} \pm \phi_{J} \pm \epsilon_- + \epsilon_+}{2}}}.
\fe
The first to the forth lines of the equations \eqref{eqn:Zplus}, \eqref{eqn:Zminusodd} and\eqref{eqn:Zminuseven} are the one-loop determinants of the ${\cal N}=4$ vector multiplet, Fermi multiplet, twisted hypermultiplet and hypermultiplet, respectively.  The one-loop determinants of the D0-D4 strings are given by
\ie\label{eqn:D0-D4Integrand}
&{\cal Z}^{+,\, k=2n+\chi}_{\text{\tiny D0-D4}}=\Big(\frac{2\sinh\tfrac{\pm\alpha-m}{2}}{2\sinh\tfrac{\pm\alpha+\epsilon_+}{2}}\Big)^\chi\prod^n_{I=1}\frac{2\sinh\tfrac{\pm\phi_I\pm\alpha-m}{2}}{2\sinh\tfrac{\pm\phi_I\pm\alpha+\epsilon_+}{2}},
\\
&{\cal Z}^{-,\,k=2n+1}_{\text{\tiny D0-D4}}=\frac{2\cosh\frac{\pm\alpha-m}{2}}{2\cosh\frac{\pm\alpha+\epsilon_+}{2}}\prod^{n}_{I=1}\frac{2\sinh\frac{\pm\phi_I\pm\alpha-m}{2}}{2\sinh\frac{\pm\phi_I\pm\alpha+\epsilon_+}{2}},
\\
&{\cal Z}^{-,\,k=2n}_{\text{\tiny D0-D4}}
=\frac{2\sinh(\pm\alpha-m)}{2\sinh(\pm\alpha+\epsilon_+)}\prod^{n-1}_{I=1}\frac{2\sinh\frac{\pm\phi_I\pm\alpha-m}{2}}{2\sinh\frac{\pm\phi_I\pm\alpha+\epsilon_+}{2}}.
\fe
The one-loop determinants of the D0-D8 strings are given by
\ie\label{eqn:D0-D8integrand}
&{\cal Z}^{+,\, k=2n+\chi}_{\text{\tiny D0-D8}}=\prod^{N_f}_{\ell=1}\Big((2\sinh\tfrac{m_\ell}{2})^\chi \prod^n_{I=1}2\sinh\tfrac{\pm\phi_I + m_\ell}{2}\Big),
\\
&{\cal Z}^{-,\,k=2n+1}_{\text{\tiny D0-D8}}=\prod^{N_f}_{\ell=1}\Big(2\cosh\tfrac{m_\ell}{2}\prod^n_{I=1}2\sinh\tfrac{\pm\phi_I + m_\ell}{2}\Big),
\\
&{\cal Z}^{-,\,k=2n}_{\text{\tiny D0-D8}}=\prod^{N_f}_{\ell=1}\Big(2\sinh m_\ell \prod^{n-1}_{I=1}2\sinh\tfrac{\pm\phi_I + m_\ell}{2}\Big).
\fe
Finally, the Weyl factors of the O$(k)_+$ and O$(k)_-$ components in \eqref{eqn:PhiIntegral}  are given by
\begin{align}
\hspace{-1cm}|W|_{+}^{\chi=0} = \frac{1}{2^{n-1} n!} ,\ |W|_{+}^{\chi=1} = \frac{1}{2^n n!} ,\ |W|^{\chi=0}_{-} =  \frac{1}{2^{n-1}(n-1)!},\ |W|^{\chi=1}_{-} = \frac{1}{2^n n!}.
\end{align}

\section{On the computation of North Pole and South Pole contributions}
\label{app:Technical}

The South Pole (and similarly North Pole) contribution to the integrands \eqref{SCI} and \eqref{eqn:ROI} is evaluated by a separate index computation of a 1d field theory (Quantum Mechanics) that describes the dynamics of strings connecting D$0$-branes to the various D-branes in the problem (D$4$-branes, D$4$'-branes, $N_f$ D$8$-branes, and the D$0$-branes themselves).
The integrals involved have been described in great detail in \cite{Hwang:2014uwa}, but for the sake of completeness we will now expand on a few of the technical details involved.

The $O(q^k)$ North Pole contribution, for $k=2n$ or $k=2n+1$, is given by an integral over $n$ variables, denoted as $\phi_1,\dots,\phi_n$.
The integrand is a fraction whose numerator and denominator are both products of terms that are contributions of individual fields of the 1d field theory, with bosonic fields contributing to the denominator and fermionic fields to the numerator. Each individual term is written as $2\sinh\SinhArg$, with $\SinhArg$ a linear expression in the equivariant parameters $\epsilon_{+}$, $\epsilon_{-}$, the $U(1)\subset\Sp(1)$ chemical potential $\alpha$, the $U(1)^{2N_f}\subset\SO(2N_f)$ chemical potentials $m_1,\dots, m_{N_f}$ and the integration variables $\phi_1,\dots,\phi_n$.
The exact form of the integrand can be found in equations (3.42)-(3.50) of \cite{Hwang:2014uwa}. We also summarized it in \appref{app:integrand}. For simplicity, we will set $m_1 = \cdots = m_{N_f}=0$ from now on.

For even $k=2n$, the integral takes the form
\be\label{eqn:IntZ2n}
{\cal Z}^{2n}_{\text{\tiny D0-D4-D$4'$-D8/O8}} = \frac{1}{2^n n!}\oint
{\cal Z}^+_{\text{\tiny D0-D0}}{\cal Z}^+_{\text{\tiny D0-D4}}{\cal Z}^+_{\text{\tiny D0-D8}}{\cal Z}^+_{\text{\tiny D0-D$4'$}}
d\phi_1\cdots d\phi_n,
\ee
where ${\cal Z}^+_{\text{\tiny D0-D0}}$, ${\cal Z}^+_{\text{\tiny D0-D4}}$, ${\cal Z}^+_{\text{\tiny D0-D8}}$ are all functions of $\phi_1,\dots,\phi_n$, $\epsilon_{+}$, $\epsilon_{-}$, and $\alpha$, and  ${\cal Z}^+_{\text{\tiny D0-D$4'$}}$  is a function of the same parameters and also $M$.
The formulas for ${\cal Z}^+_{\text{\tiny D0-D0}}$, ${\cal Z}^+_{\text{\tiny D0-D4}}$, ${\cal Z}^+_{\text{\tiny D0-D8}}$ are 
\ie
&{\cal Z}_{\text{\tiny D0-D0}}^{+}=  \bigg( {2 \sinh{\tfrac{\pm m - \epsilon_-}{2}}\over 2\sinh{\frac{\pm m - \epsilon_+}{2}}2 \sinh{\frac{\pm \epsilon_- + \epsilon_+ }{2}}}\bigg)^{n}\prod_{I=1}^{n} \frac{1}{2 \sinh{\frac{\pm 2\phi_I \pm m - \epsilon_+}{2}}   2\sinh{ \frac{\pm 2\phi_{I} \pm \epsilon_- + \epsilon_+}{2}}} 
\\
&~~~~~~~~~~~~\times\prod_{I<J}^{n} \frac{2\sinh{ \tfrac{ \pm \phi_I \pm \phi_J}{2}}2\sinh{ \tfrac{\pm \phi_{I} \pm \phi_{J} + 2\epsilon_+}{2} }2\sinh{\tfrac{\pm\phi_I \pm \phi_J \pm m - \epsilon_-}{2}}}{2\sinh{\frac{\pm\phi_I \pm \phi_J \pm m - \epsilon_+}{2}}2\sinh{ \frac{\pm \phi_{I} \pm \phi_{J} \pm \epsilon_- + \epsilon_+}{2}}},
\\
&{\cal Z}^{+}_{\text{\tiny D0-D4}}=\prod^n_{I=1}\frac{2\sinh\tfrac{\pm\phi_I\pm\alpha-m}{2}}{2\sinh\tfrac{\pm\phi_I\pm\alpha+\epsilon_+}{2}},~~~{\cal Z}^{+}_{\text{\tiny D0-D8}}=\prod^{N_f}_{\ell=1}\prod^n_{I=1}2\sinh\tfrac{\pm\phi_I + m_\ell}{2}.\nn
\fe
The additional parameter $m$ that appears in ${\cal Z}^+_{\text{\tiny D0-D0}}$ and ${\cal Z}^+_{\text{\tiny D0-D4}}$ represents an additional twist that can be set to $m=0$, but is kept nonzero in intermediate stages of the computation in order to regularize the integral over $\phi_1,\dots,\phi_n$, as we shall review below.
The formulas for odd $k$ are of a similar spirit, but slightly more complicated, and can be found in \cite{Hwang:2014uwa}, and also copied in \appref{app:integrand}. The formula for ${\cal Z}^{+}_{\text{\tiny D0-D$4'$}}$ is
\be\label{eqn:ZprimePlus}
{\cal Z}^{+}_{\text{\tiny D0-D$4'$}} = \prod_{I=1}^n\frac{
2\sinh\frac{\pm\phi_I\pm M-\eM}{2}
}{
2\sinh\frac{\pm\phi_I\pm M-\eP}{2}
}\,.
\ee

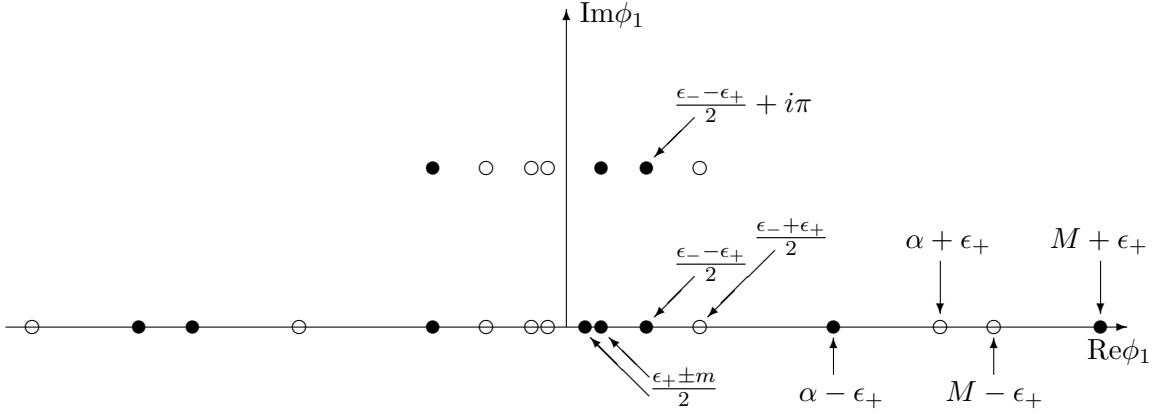
\begin{figure}[t]
\begin{picture}(420,170)
\put(210,50){\begin{picture}(0,0)
\put(-210,0){\vector(1,0){420}}
\put(0,0){\vector(0,1){120}}
\put(195,-12){$\text{Re}\phi_1$}
\put(5,115){$\text{Im}\phi_1$}

\put(-50,0){\circle*{5}}
\put(-50,60){\circle*{5}}
\put(30,0){\circle*{5}}\put(40,21){$\tfrac{\eM-\eP}{2}$}\put(48,19){\vector(-1,-1){15}}
\put(30,60){\circle*{5}}\put(40,81){$\tfrac{\eM-\eP}{2}+i\pi$}\put(48,79){\vector(-1,-1){15}}
\put(100,0){\circle*{5}}\put(87,-28){$\alpha-\eP$}\put(100,-18){\vector(0,1){14}}

\put(-140,0){\circle*{5}}
\put(13,0){\circle*{5}}\put(31,-26){$\tfrac{\eP\pm m}{2}$}
\put(31,-19){\vector(-1,1){15}}
\put(31,-26){\vector(-1,1){22}}
\put(7,0){\circle*{5}}
\put(13,60){\circle*{5}}
\put(200,0){\circle*{5}}\put(181,30){$M+\eP$}\put(200,25){\vector(0,-1){20}}
\put(-160,0){\circle*{5}}

\put(50,0){\circle{5}}\put(70,31){$\tfrac{\eM+\eP}{2}$}\put(78,29){\vector(-1,-1){25}}

\put(50,60){\circle{5}}
\put(-30,0){\circle{5}}
\put(-30,60){\circle{5}}
\put(-100,0){\circle{5}}
\put(140,0){\circle{5}}\put(127,30){$\alpha+\eP$}\put(140,25){\vector(0,-1){20}}

\put(-13,0){\circle{5}}
\put(-7,0){\circle{5}}
\put(-13,60){\circle{5}}
\put(-7,60){\circle{5}}
\put(160,0){\circle{5}}\put(142,-28){$M-\eP$}\put(160,-18){\vector(0,1){14}}
\put(-200,0){\circle{5}}

\end{picture}}
\end{picture}
\caption{
The location of the poles on the complex $\phi_1$ plane for instanton number $k=2$. The filled circles indicate the poles that are retained by the Jeffrey-Kirwan prescription, while the hollow circles indicate the poles that are ignored.
}
\label{fig:JKpoles1}
\end{figure}

The integration parameters $\phi_I$ ($I=1,\dots,n$) live on a cylinder, with $-\infty<\text{Re}\phi_I<\infty$, and $0\le\text{Im}\phi_I<2\pi$ periodic.
The integral \eqref{eqn:IntZ2n} is performed by summing over the contributions of the poles within the integration path, which we have not described yet. A pole can arise when an argument of a sinh in the denominator of ${\cal Z}^{+}_{\text{\tiny D0-D0}}$ or ${\cal Z}^{+}_{\text{\tiny D0-D4}}$ equals a multiple of $\pi i$.
Which poles to keep was determined in  \cite{Hwang:2014uwa}, using the Jeffrey-Kirwan (JK) residue technique developed in \cite{Jeffrey:aa} and explained in \cite{Benini:2013nda,Benini:2013xpa}.
For $n=1$ the integration is one-dimensional and the JK prescription is to consider only the poles arising from the terms where the coefficient of $\phi_1$ in the argument of sinh is positive. These are $14$ poles, which we list below:
\be\label{eqn:n=1JKPoles}
\phi_1\rightarrow
\pm\tfrac{1}{2}\eM-\tfrac{1}{2}\eP,
\quad
\pm\tfrac{1}{2}\eM-\tfrac{1}{2}\eP + i\pi,
\quad
\pm\alpha-\eP,
\quad
\pm\tfrac{1}{2}m+\tfrac{1}{2}\eP,
\quad
\pm\tfrac{1}{2}m+\tfrac{1}{2}\eP+ i\pi,
\ee
and
$$
\phi_1\rightarrow\pm M+\eP.
$$
For generic $M$, $m$, $\eM$, $\eP$, and $\alpha$, these are all simple poles, but when we set $m\rightarrow 0$, we get a double pole at $\phi_1=\tfrac{1}{2}\eP$. For $n=1$, keeping $m\neq 0$ in intermediate steps is only a convenience. It will become crucial for $n>1$.
The poles are depicted in \figref{fig:JKpoles1} in the regime
$m\ll\eP\ll\eM\ll\alpha$.

For $k=3$ the index is similarly calculated by an integral over a single parameter $\phi_1$, but for $k=4$ the integral is over two parameters $d\phi_1 d\phi_2$, and the prescription is as follows.
Residues of poles are evaluated at values of $(\phi_1,\phi_2)$ where the arguments of at least two different sinh's in the denominator of the integrand are an integer multiple of $i\pi$. They are a simple pole if exactly two sinh's vanish. The argument of the $i^{th}$ sinh ($i=1,2$) takes the form $\sum_I\JKQ_{iI}\phi_I+\zeta_i$, where $\JKQ_{iI}$ are constants (taking the possible values $0$, $\pm 1/2$ or $\pm 1$), and $\zeta_i$ are independent of $\phi_1$ and $\phi_2$ (and are linear  expressions in $\eP$, $\eM$, $m$, $M$, $\alpha$). The Jeffrey-Kirwan prescription requires us to fix an arbitrary (row) vector $\JKeta\equiv(\JKeta_1\, \JKeta_2)$, then calculate, for each pole, the vector $\JKeta\JKQ^{-1}$, and keep the residue only if all the components of $\JKeta\JKQ^{-1}$ are positive. (In other words, $\eta$ has to be inside the cone generated by the rows of $\JKQ$.)

Double poles appear at
$$
\phi_1 = \pm\phi_2 = \pm\tfrac{1}{2}m\pm\tfrac{1}{2}\eP
\qquad\text{($\pm$ signs are uncorrelated),}
$$
where also one of the expressions $\pm\phi_1\pm\phi_2\pm m-\eP$
(for the appropriate sign assignments) vanishes. These are $8$ in number, and there are additional $8$ double poles at
$$
\phi_1 = \pm\phi_2 = \pm\tfrac{1}{2}\eM\pm\tfrac{1}{2}\eP
\qquad\text{($\pm$ signs are uncorrelated),}
$$
where also one of the expressions $(\pm\phi_1\pm\phi_2\pm\eM+\eP)/2$ vanishes.
However, in these cases also $\pm\phi_1\pm\phi_2$ vanishes (for two sign assignments), which gives the numerator of ${\cal Z}^+_{\text{\tiny D0-D0}}$ a double zero, and these poles therefore do not contribute to the integral. In the above discussion, we can also add $i\pi$ to both $\phi_1$ and $\phi_2$ and get another set of eight double poles, but if we add $i\pi$ to only $\phi_1$ or only $\phi_2$, we get a simple pole.
Ignoring the above mentioned double poles, for $k=4$ there are $352$  simple poles that pass the Jeffrey-Kirwan requirement.

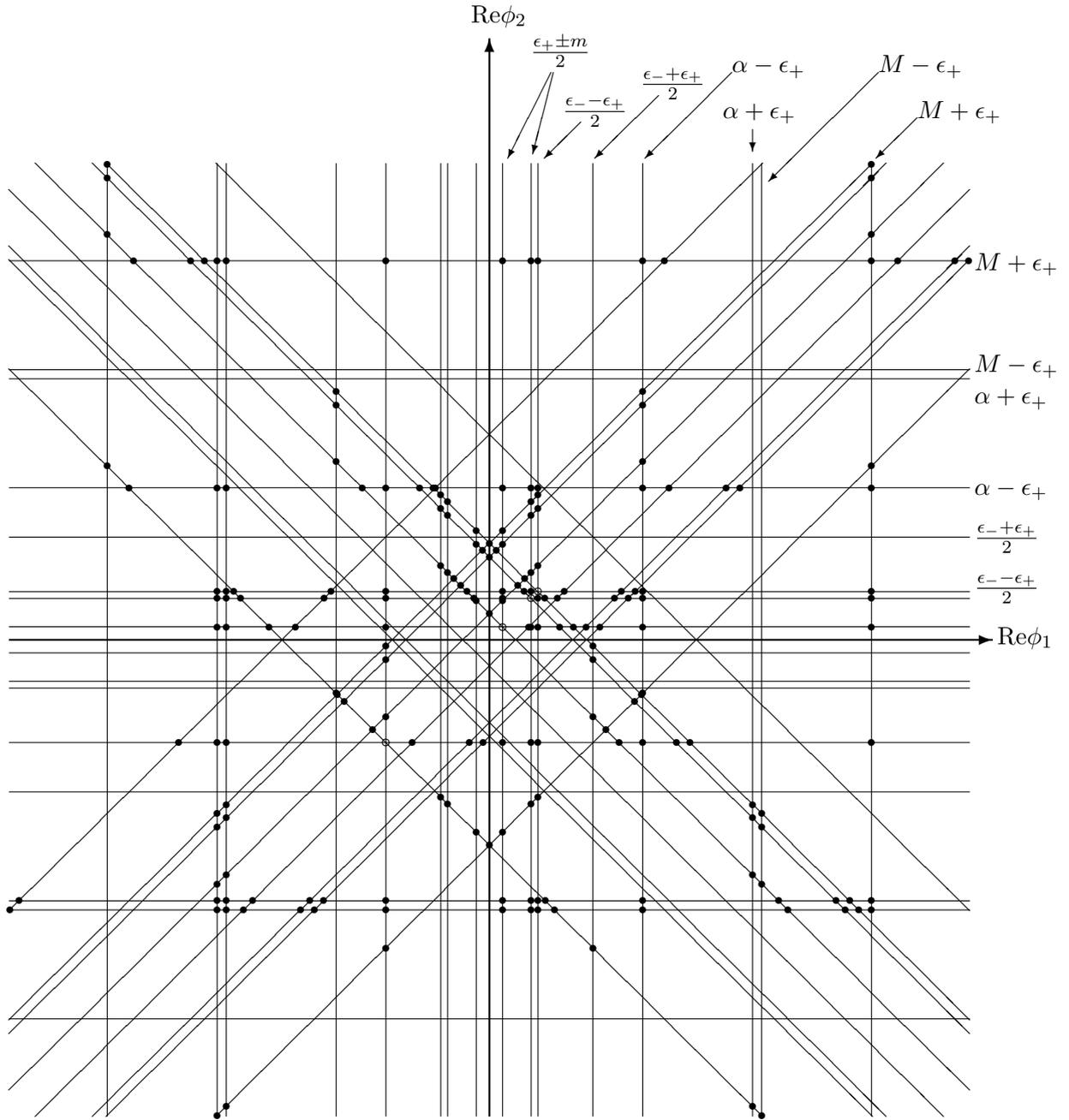
\begin{figure}[t]
\begin{picture}(440,490)
\put(210,210){\begin{picture}(0,0)
\thicklines
\put(222,-2){$\text{Re}\phi_1$}
\put(-8,272){$\text{Re}\phi_2$}

\put(0,-210){\vector(0,1){475}}
\put(-210,0){\vector(1,0){430}}

\thinlines
\put(-210.,119.5){\line(1,-1){329.5}}
\put(-119.5,-210.){\line(1,1){329.5}}
\put(-210.,-119.5){\line(1,1){329.5}}
\put(-119.5,210.){\line(1,-1){329.5}}
\put(-167.5,210.){\line(1,-1){377.5}}
\put(-210.,-167.5){\line(1,1){377.5}}
\put(-167.5,-210.){\line(1,1){377.5}}
\put(-210.,167.5){\line(1,-1){377.5}}
\put(-45.25,-210.){\line(0,1){420.}}
\put(45.25,-210.){\line(0,1){420.}}
\put(21.25,-210.){\line(0,1){420.}}
\put(-21.25,-210.){\line(0,1){420.}}
\put(-210.,-45.25){\line(1,0){420.}}
\put(-210.,45.25){\line(1,0){420.}}
\put(-210.,21.25){\line(1,0){420.}}
\put(-210.,-21.25){\line(1,0){420.}}
\put(-115,-210){\line(0,1){420}}
\put(115,-210){\line(0,1){420}}
\put(67,-210){\line(0,1){420}}
\put(-67,-210){\line(0,1){420}}
\put(-210,-115){\line(1,0){420}}
\put(-210,115){\line(1,0){420}}
\put(-210,67){\line(1,0){420}}
\put(-210,-67){\line(1,0){420}}
\put(-119,-210){\line(0,1){420}}
\put(119,-210){\line(0,1){420}}
\put(167,-210){\line(0,1){420}}
\put(-167,-210){\line(0,1){420}}
\put(-210,-119){\line(1,0){420}}
\put(-210,119){\line(1,0){420}}
\put(-210,167){\line(1,0){420}}
\put(-210,-167){\line(1,0){420}}
\put(-198.5,210.){\line(1,-1){408.5}}
\put(-210.,-198.5){\line(1,1){408.5}}
\put(-198.5,-210.){\line(1,1){408.5}}
\put(-210.,198.5){\line(1,-1){408.5}}
\put(-173.5,210.){\line(1,-1){383.5}}
\put(-210.,-173.5){\line(1,1){383.5}}
\put(-173.5,-210.){\line(1,1){383.5}}
\put(-210.,173.5){\line(1,-1){383.5}}
\put(5.75,-210.){\line(0,1){420.}}
\put(-5.75,-210.){\line(0,1){420.}}
\put(18.25,-210.){\line(0,1){420.}}
\put(-18.25,-210.){\line(0,1){420.}}
\put(-210.,5.75){\line(1,0){420.}}
\put(-210.,-5.75){\line(1,0){420.}}
\put(-210.,18.25){\line(1,0){420.}}
\put(-210.,-18.25){\line(1,0){420.}}
\thinlines
\put(0,-90.5){\circle*{3}}
\put(-66.5,-24){\circle*{3}}
\put(45.25,-135.75){\circle*{3}}
\put(-21.25,-69.25){\circle*{3}}
\put(-45.25,-45.25){\circle{3}}
\put(-111.75,21.25){\circle*{3}}
\put(115,-205.5){\circle*{3}}
\put(-67,-23.5){\circle*{3}}
\put(24.5,-115){\circle*{3}}
\put(-157.5,67){\circle*{3}}
\put(119,-209.5){\circle*{3}}
\put(-167,76.5){\circle*{3}}
\put(28.5,-119){\circle*{3}}
\put(-51.,-39.5){\circle*{3}}
\put(-63.5,-27.){\circle*{3}}
\put(-5.75,-84.75){\circle*{3}}
\put(-18.25,-72.25){\circle*{3}}
\put(-96.25,5.75){\circle*{3}}
\put(-108.75,18.25){\circle*{3}}
\put(66.5,-24){\circle*{3}}
\put(-45.25,-135.75){\circle*{3}}
\put(21.25,-69.25){\circle*{3}}
\put(-115,-205.5){\circle*{3}}
\put(67,-23.5){\circle*{3}}
\put(-119,-209.5){\circle*{3}}
\put(167,76.5){\circle*{3}}
\put(51.,-39.5){\circle*{3}}
\put(63.5,-27.){\circle*{3}}
\put(5.75,-84.75){\circle*{3}}
\put(18.25,-72.25){\circle*{3}}
\put(-135.75,-45.25){\circle*{3}}
\put(-69.25,21.25){\circle*{3}}
\put(-205.5,-115){\circle*{3}}
\put(-23.5,67){\circle*{3}}
\put(-209.5,-119){\circle*{3}}
\put(76.5,167){\circle*{3}}
\put(-84.75,5.75){\circle*{3}}
\put(-72.25,18.25){\circle*{3}}
\put(0,42.5){\circle*{3}}
\put(45.25,-2.75){\circle*{3}}
\put(-21.25,63.75){\circle*{3}}
\put(87.75,-45.25){\circle*{3}}
\put(21.25,21.25){\circle{3}}
\put(115,-72.5){\circle*{3}}
\put(-67,109.5){\circle*{3}}
\put(157.5,-115){\circle*{3}}
\put(-24.5,67){\circle*{3}}
\put(119,-76.5){\circle*{3}}
\put(-167,209.5){\circle*{3}}
\put(161.5,-119){\circle*{3}}
\put(-124.5,167){\circle*{3}}
\put(15.5,27.){\circle*{3}}
\put(3.,39.5){\circle*{3}}
\put(-5.75,48.25){\circle*{3}}
\put(-18.25,60.75){\circle*{3}}
\put(36.75,5.75){\circle*{3}}
\put(24.25,18.25){\circle*{3}}
\put(-45.25,-2.75){\circle*{3}}
\put(21.25,63.75){\circle*{3}}
\put(-115,-72.5){\circle*{3}}
\put(67,109.5){\circle*{3}}
\put(-119,-76.5){\circle*{3}}
\put(167,209.5){\circle*{3}}
\put(-15.5,27.){\circle*{3}}
\put(-3.,39.5){\circle*{3}}
\put(5.75,48.25){\circle*{3}}
\put(18.25,60.75){\circle*{3}}
\put(-2.75,-45.25){\circle*{3}}
\put(63.75,21.25){\circle*{3}}
\put(-72.5,-115){\circle*{3}}
\put(109.5,67){\circle*{3}}
\put(-76.5,-119){\circle*{3}}
\put(209.5,167){\circle*{3}}
\put(48.25,5.75){\circle*{3}}
\put(60.75,18.25){\circle*{3}}
\put(-45.25,21.25){\circle*{3}}
\put(-45.25,-115){\circle*{3}}
\put(-45.25,67){\circle*{3}}
\put(-45.25,-119){\circle*{3}}
\put(-45.25,167){\circle*{3}}
\put(-45.25,-33.75){\circle*{3}}
\put(-45.25,-8.75){\circle*{3}}
\put(-45.25,5.75){\circle*{3}}
\put(-45.25,18.25){\circle*{3}}
\put(45.25,-33.75){\circle*{3}}
\put(45.25,-8.75){\circle*{3}}
\put(21.25,-45.25){\circle*{3}}
\put(21.25,-115){\circle*{3}}
\put(21.25,67){\circle*{3}}
\put(21.25,-119){\circle*{3}}
\put(21.25,167){\circle*{3}}
\put(21.25,32.75){\circle*{3}}
\put(21.25,57.75){\circle*{3}}
\put(21.25,5.75){\circle*{3}}
\put(21.25,18.25){\circle*{3}}
\put(-21.25,32.75){\circle*{3}}
\put(-21.25,57.75){\circle*{3}}
\put(-115,-45.25){\circle*{3}}
\put(67,-45.25){\circle*{3}}
\put(-119,-45.25){\circle*{3}}
\put(167,-45.25){\circle*{3}}
\put(56.75,-45.25){\circle*{3}}
\put(-33.75,-45.25){\circle*{3}}
\put(81.75,-45.25){\circle*{3}}
\put(-8.75,-45.25){\circle*{3}}
\put(5.75,-45.25){\circle*{3}}
\put(18.25,-45.25){\circle*{3}}
\put(-115,21.25){\circle*{3}}
\put(67,21.25){\circle*{3}}
\put(-119,21.25){\circle*{3}}
\put(167,21.25){\circle*{3}}
\put(-9.75,21.25){\circle*{3}}
\put(32.75,21.25){\circle*{3}}
\put(15.25,21.25){\circle*{3}}
\put(57.75,21.25){\circle*{3}}
\put(5.75,21.25){\circle*{3}}
\put(18.25,21.25){\circle*{3}}
\put(-115,-115){\circle*{3}}
\put(-115,67){\circle*{3}}
\put(-115,-119){\circle*{3}}
\put(-115,167){\circle*{3}}
\put(-115,-103.5){\circle*{3}}
\put(-115,-78.5){\circle*{3}}
\put(-115,5.75){\circle*{3}}
\put(-115,18.25){\circle*{3}}
\put(115,-103.5){\circle*{3}}
\put(115,-78.5){\circle*{3}}
\put(67,-115){\circle*{3}}
\put(67,67){\circle*{3}}
\put(67,-119){\circle*{3}}
\put(67,167){\circle*{3}}
\put(67,78.5){\circle*{3}}
\put(67,103.5){\circle*{3}}
\put(67,5.75){\circle*{3}}
\put(67,18.25){\circle*{3}}
\put(-67,78.5){\circle*{3}}
\put(-67,103.5){\circle*{3}}
\put(-119,-115){\circle*{3}}
\put(167,-115){\circle*{3}}
\put(126.5,-115){\circle*{3}}
\put(-103.5,-115){\circle*{3}}
\put(151.5,-115){\circle*{3}}
\put(-78.5,-115){\circle*{3}}
\put(5.75,-115){\circle*{3}}
\put(18.25,-115){\circle*{3}}
\put(-119,67){\circle*{3}}
\put(167,67){\circle*{3}}
\put(-55.5,67){\circle*{3}}
\put(78.5,67){\circle*{3}}
\put(-30.5,67){\circle*{3}}
\put(103.5,67){\circle*{3}}
\put(5.75,67){\circle*{3}}
\put(18.25,67){\circle*{3}}
\put(-119,-119){\circle*{3}}
\put(-119,167){\circle*{3}}
\put(-119,-107.5){\circle*{3}}
\put(-119,-82.5){\circle*{3}}
\put(-119,5.75){\circle*{3}}
\put(-119,18.25){\circle*{3}}
\put(119,-107.5){\circle*{3}}
\put(119,-82.5){\circle*{3}}
\put(167,-119){\circle*{3}}
\put(167,167){\circle*{3}}
\put(167,178.5){\circle*{3}}
\put(167,203.5){\circle*{3}}
\put(167,5.75){\circle*{3}}
\put(167,18.25){\circle*{3}}
\put(-167,178.5){\circle*{3}}
\put(-167,203.5){\circle*{3}}
\put(130.5,-119){\circle*{3}}
\put(-107.5,-119){\circle*{3}}
\put(155.5,-119){\circle*{3}}
\put(-82.5,-119){\circle*{3}}
\put(5.75,-119){\circle*{3}}
\put(18.25,-119){\circle*{3}}
\put(-155.5,167){\circle*{3}}
\put(178.5,167){\circle*{3}}
\put(-130.5,167){\circle*{3}}
\put(203.5,167){\circle*{3}}
\put(5.75,167){\circle*{3}}
\put(18.25,167){\circle*{3}}
\put(0,11.5){\circle*{3}}
\put(-12.5,24){\circle*{3}}
\put(-5.75,17.25){\circle*{3}}
\put(-18.25,29.75){\circle*{3}}
\put(5.75,5.75){\circle{3}}
\put(-6.75,18.25){\circle*{3}}
\put(12.5,24){\circle*{3}}
\put(5.75,17.25){\circle*{3}}
\put(18.25,29.75){\circle*{3}}
\put(17.25,5.75){\circle*{3}}
\put(29.75,18.25){\circle*{3}}
\put(0,36.5){\circle*{3}}
\put(-5.75,42.25){\circle*{3}}
\put(-18.25,54.75){\circle*{3}}
\put(30.75,5.75){\circle*{3}}
\put(18.25,18.25){\circle{3}}
\put(5.75,42.25){\circle*{3}}
\put(18.25,54.75){\circle*{3}}
\put(42.25,5.75){\circle*{3}}
\put(54.75,18.25){\circle*{3}}
\put(5.75,18.25){\circle*{3}}
\put(18.25,5.75){\circle*{3}}


\put(0,210){\begin{picture}(0,0)
\put(32,20){$\tfrac{\eM-\eP}{2}$}\put(40,18){\vector(-1,-1){16}}
\put(102,20){$\alpha+\eP$}\put(115,15){\vector(0,-1){10}}
\put(106,40){$\alpha-\eP$}\put(106,40){\vector(-1,-1){38}}
\put(18,46){$\tfrac{\eP\pm m}{2}$}
\put(27,40){\vector(-1,-2){19}}
\put(28,40){\vector(-1,-4){9}}
\put(66,32){$\tfrac{\eM+\eP}{2}$}\put(74,30){\vector(-1,-1){28}}
\put(170,40){$M-\eP$}\put(170,40){\vector(-1,-1){48}}
\put(187,20){$M+\eP$}\put(187,20){\vector(-1,-1){18}}
\end{picture}}

\put(212,-2){\begin{picture}(0,0)
\put(0,22){$\tfrac{\eM-\eP}{2}$}
\put(0,165){$M+\eP$}
\put(0,120){$M-\eP$}
\put(0,106){$\alpha+\eP$}
\put(0,66){$\alpha-\eP$}
\put(0,44){$\tfrac{\eM+\eP}{2}$}
\end{picture}}

\end{picture}}
\end{picture}
\caption{
The location of the poles on the $\phi_1-\phi_2$ real plane for instanton number $k=4$, for $\eta=(1,3)$. 
The lines are the loci where the argument of a single sinh in the denominator of the integrand vanishes. Poles are at the intersection of two lines.
 The solid circles indicate the poles that are retained by the Jeffrey-Kirwan prescription.
(One pole, at $\phi_1=-M-\eM-2\eP$ and $\phi_2=M+\eP$, is outside the frame of the picture.)
The hollow circles are possible locations of non-simple poles, where three lines intersect.
(Whether they are simple or non-simple depends on $\text{Im}\phi_1$ and $\text{Im}\phi_2$.)
}
\label{fig:JKpoles2}
\end{figure}

\section{D-branes in massive type IIA}
\label{app:miia}
The massive type IIA supergravity action is given by\footnote{The action is invariant under the NSNS gauge transformation, where the usual $B_2$-field transformation $\delta B_2=d\Lambda_1$ is accompanied with the transformation of the RR-fields $\delta C_1=-\mIIAM\Lambda_1$ and $\delta C_3=\mIIAM\Lambda_1\wedge B_2$.
}
\ie\label{eqn:SGaction}
&{\bf S}_{\text{\tiny NS}}={1\over 2\kappa^2_{10}}\int d^{10} x\sqrt{-G}\,e^{-2\Phi}\left(R+4\partial_\mu\Phi\partial^\mu\Phi - {1\over 2}|H_3|^2\right),
\\
&{\bf S}_{\text{\tiny R}}=-{1\over 4\kappa^2_{10}}\int d^{10} x\sqrt{-G}\left(|F_2+ \mIIAM B_2|^2+|\widetilde F_4-{1\over 2}\mIIAM B_2^2|^2\right),
\\
&{\bf S}_{\text{\tiny CS}}=-{1\over 4\kappa^2_{10}}\int \left\{B_2\wedge F_4^2-{1\over 3}\mIIAM B_2^3\wedge F_4 +{1\over 20}\mIIAM^2 B_2^5\right\},
\\
&{\bf S}_{\text{\tiny mass}}=-{1\over 4\kappa_{10}^2}\int d^{10}x\sqrt{-G}\,\mIIAM^2+{1\over 2\kappa^2_{10}}\int  \mIIAM F_{10},
\fe
where the $\widetilde F_4$ is defined by
\ie
\widetilde F_4=dC_3-C_1\wedge dB_2.
\fe
Consider a D8-brane localized at a constant value of $x^9$, say at $x^9=0$. It behaves like a domain wall that splits the spacetime into two regions $x^9<0$ and $x^9>0$. The action of the D8-brane is given by
\ie
{\bf S}_{\text{\tiny D8}}=-\mu_8\int d^9x\,e^{-\Phi} \sqrt{-G^{(9)}} + \mu_8\int C_9.
\fe
In this appendix, we use Polchinski's convention \cite{Polchinski:1998rr}. The gravitational coupling $\kappa_{10}$ and the D$p$-brane charge $\mu_p$ are given by
\ie
\kappa_{10}^2={1\over 2}(2\pi)^7\alpha'^4,~~~\mu_p^2= (2\pi)^{-2p}\alpha^{-p-1}.
\fe 

Varying the total action by $C_9$ gives the equation of motion of the Romans mass $\mIIAM$,
\ie\label{Mjump}
{\partial \mIIAM\over \partial x^9}=2\kappa_{10}^2\mu_8\delta(D8),
\fe
which implies that the Romans mass jumps by $2\kappa_{10}^2\mu_8$ when crossing the D8-brane. Similarly, the derivative of the dilaton jumps when crossing the D8-brane\footnote{The simplest way to derive this relation is to consider the equation of motion of the dilaton in the Einstein frame $G^{\rm E}_{\mu\nu}=e^{-{1\over 2}\Phi}G_{\mu\nu}$,
\ie
\nabla^{\rm E}_\mu\partial^\mu\Phi-{5\over 4}\mIIAM^2 e^{{5\over 2}\Phi}={5\over 2}\mu_8\kappa_{10}^2(G^E_{99})^{-{1\over 2}}e^{{5\over 4}\Phi}\delta(D8).
\fe
}
\ie\label{dPhiJump}
\partial_9 \Phi\Big|_{x^9=0^+}-\partial_9 \Phi\Big|_{x^9=0^-}={5\over 2}\mu_8 \kappa_{10}^2 e^{\Phi(0)} \sqrt{G_{99}(0)}.
\fe

Away from the D8-brane, the equations of motion of the dilaton $\Phi$ and the metric $G_{\mu\nu}$, with all the other fields setting to zero, are given by
\ie
&R_{\mu\nu}+2\nabla_\mu \partial_\nu \Phi-{1\over 2}G_{\mu\nu}\left(R+4\nabla^\rho\partial_\rho\Phi-4\partial^\rho\Phi\partial_\rho\Phi-{1\over 2}\mIIAM^2e^{2\Phi}\right)=0
\\
&R+4\nabla^\mu\partial_\mu\Phi-4\partial^\mu\Phi\partial_\mu\Phi=0
\fe
Let us consider a domain wall ansatz,
\ie
ds^2=\Omega^2(x^9) \eta_{\mu\nu}dx^\mu dx^\nu,~~~\Phi=\Phi(x^9).
\fe 
The solution to the equations is given by
\ie\label{DMS}
&\Omega(x^9)={2\over 3}c_2 (c_1\pm c_2 \mIIAM x^9)^{-{1\over 6}},~~~e^{\Phi(x^9)}=(c_1\pm c_2 \mIIAM x^9)^{-{5\over 6}},
\fe
$c_1$ and $c_2$ are constant away from the D8-brane. By the equations \eqref{Mjump} and \eqref{dPhiJump} and the continuity of the metric and dilaton, $c_1$ and $c_2$ still remain constant when crossing the D8-brane, and we must take the lower sign in \eqref{DMS}. By a coordinate transformation $x'^9= {1\over \mIIAM}(c_1-c_2 \mIIAM x^9)^{{2\over 3}}$, the solution can be put into the form as (relabel $x'^9$ by $x^9$)
\ie\label{eqn:miiAbackground}
e^{\Phi(x^9)}=(\mIIAM x^9)^{-{5\over 4}},~~ds^2= (\mIIAM x^9)^{-{1\over 2}}\left[-(dx^0)^2+(dx^1)^2+\cdots+(dx^8)^2\right]+(\mIIAM x^9)^{{1\over 2}}(dx^9)^2.
\fe

Now, let us focus on the case of interest: $N_f$ D8-branes coinciding with O8 plane in the strong coupling limit. The D8/O8 singularity is located at $x^9=0$, where the string coupling diverges. The total RR 9-form charge is $\mFlux\equiv(8-N_f)$, and the Romans mass is given by
\ie
\mIIAM=2\mFlux\mu_8\kappa_{10}^2={\mFlux\over 2\pi \sqrt{\alpha'}}.
\fe
We introduce a D4-brane located at $y>0$. The DBI action of the U(1) gauge theory on the D4-brane worldvolume is given by
\ie
{\bf S}_{\rm DBI}&=-\mu_4\int d^5 x\,e^{-\Phi}\left[-\det(G^{(5)}_{ab}+B_{ab}+2\pi \alpha' {\bf f}_{ab})\right]^{1/2}.
\fe
In the static gauge, the induced metric $G^{(5)}_{ab}$ is given by
\ie
G^{(5)}_{ab}=(\mIIAM x^9)^{-{1\over 2}}\left(\eta_{ab} + (2\pi \alpha')^2 \delta_{AB}\partial_a X^A \partial_b X^B\right)+ (2\pi \alpha')^2  (\mIIAM x^9)^{{1\over 2}}\partial_a \varphi\partial _b \varphi,
\fe
where $a,b=0,1,\cdots,4$ and $A,B=5,\cdots,8$. We expand the DBI action
\ie\label{eqn:DBIaction}
{\bf S}_{\text{\tiny DBI}}&=-\mu_4\,{\bf vol}_{D4}-{1\over 2g_{\rm ym}^2(v)}\int d^5 x\, \Big[{1\over 2}\big( {\bf f}+{1\over 2\pi \alpha'}B_2\big)_{ab}\big({\bf f}+{1\over 2\pi \alpha'}B_2\big)^{ab} + \eta^{ab}\partial_a \varphi\partial _b\varphi\Big]
\\
&~~~~~~~~-{1\over 8\pi^2\sqrt{\alpha'}}\int d^5 x\, \eta^{ab}\delta_{AB}\partial_a X^A\partial _bX^B+{\cal O}(\alpha^3),
\fe
where $*_5$ is the Hodge star operator with respect to the 5-dimensional flat metric. The Yang-Mills coupling $g_{\rm ym}$ is determined by vev of the scalar field $v= \vev{\varphi}=x^9/2\pi \alpha'$ as
\ie
{1\over g_{\rm ym}^2(v)}=\mu_4(2\pi \alpha')^2\mIIAM  x^9={\mFlux v\over 4\pi^2}.
\fe

The Wess-Zumino action on D4-brane worldvolume is given by
\ie\label{eqn:WZaction}
{\bf S}_{\text{\tiny WZ}}
&=\mu_4\left[\int C_5+\int (2\pi \alpha' {\bf f}+B_2)\wedge C_3+{1\over 2}\int (2\pi \alpha' {\bf f}+B_2)^2\wedge C_1\right].
\fe
There is an additional Chern-Simons term \cite{Green:1996bh}
\ie
{\bf S}_{\text{\tiny CS}}=-{1\over 6}\mu_4 \mIIAM (2\pi \alpha' )^3 \int   {\bf a}\wedge {\bf f}^2,
\fe
which is required to maintain gauge invariance under the NSNS gauge transformation,
\ie
\delta B_2=d\Lambda_1,~~{\bf a}=-{1\over 2\pi \alpha'}\Lambda_1,~~\delta C_1=-\mIIAM\Lambda_1,~~\delta C_3=\mIIAM\Lambda_1\wedge B_2,~~\delta C_5=-{1\over 2}\mIIAM\Lambda_1\wedge B_2^2.
\fe
In general, the Chern-Simons action on the D$p$-brane worldvolume reads
\ie
{\bf S}_{\text{\tiny CS}} =  - {1\over \left({p\over 2}+1\right)!}\mu_p \mIIAM (2\pi \alpha' )^{{p\over 2}+1}\int_{p+1}{\bf a}\wedge {\bf f}^{{p\over 2}} = - {1\over \left({p\over 2}+1\right)!(2\pi)^{p\over 2}}\mFlux\int_{p+1}{\bf a}\wedge {\bf f}^{{p\over 2}} .
\fe

\section{$E_2$ group theory}
\label{app:E2}

The case $N_f=1$ corresponds to flavor group $E_2\cong$ SU(2)$\times$U(1).
In \eqref{eqn:yzqy1} we used a relation, given in (4.10) of \cite{Kim:2012gu}, to convert the fugacities associated with $E_2$ to fugacities associated with the SO$(2N_f)\times $U(1)${}_I$ subgroup that is manifest in the index formula [and we added the subscript $I$ to distinguish U(1)${}_I$ from the U(1) factor of $E_2$]. We will now explain the origin of \eqref{eqn:yzqy1}. Pick a Cartan subalgebra U$(1)'\subset$ SU(2), and consider a state with U$(1)'\times$U(1) $\subset E_2$ charges $Q'$ and $\widetilde{Q}$.
With the fugacities defined in \eqref{eqn:yzqy1}, its contribution to the index is $y^{Q'}z^{\widetilde{Q}}$, which can also be written as $y_1^{Q_1}q^{Q_I}$, where $Q_1$ is the charge associated with SO$(2N_f)=$ SO(2) $\cong$ U(1) and $Q_I$ is the instanton charge associated with U(1)${}_I$.
According to \eqref{eqn:yzqy1}, the charges are related by
\be\label{eqn:Q1QIdict}
Q_1=\tfrac{1}{2}Q'+\tfrac{7}{2}\widetilde{Q}
\,,\qquad
Q_I=\tfrac{1}{2}Q'-\tfrac{1}{2}\widetilde{Q}.
\ee
These relations have a nice string theory interpretation in terms of the D$0$-D$8$/O$8$ system, following the analysis of \cite{Danielsson:1996es,Kachru:1996nd,Lowe:1997fc,Aharony:1997pm}.
Consider an O$8$ plane with $N_f=1$ D$8$-brane, and separate the D$8$-brane from the orientifold plane.
The W-boson of SU(2) $\subset$ SU(2)$\times$U(1) $\cong E_2$ can be constructed as an open fundamental string connecting the D$8$-brane to a D$0$-brane that is stuck on the O$8$-plane. This string has charges $Q_I = Q_1 =1$, and since it is the W-boson of SU(2), it has charge $Q'=2$ and $\widetilde{Q}=0$, which is consistent with \eqref{eqn:Q1QIdict}. [Our normalization has charge $Q'=\pm 1$ for the fundamental representation $\rep{2}$ of SU(2).]
On the other hand, we can construct an SU(2) neutral state from a D$0$-brane connected by $8-N_f=7$ strings to the D$8$-brane. This particle has charges $Q_I=1$, $Q_1=-7$, $Q'=0$ and $\widetilde{Q}=-2$, again consistent with \eqref{eqn:Q1QIdict}.

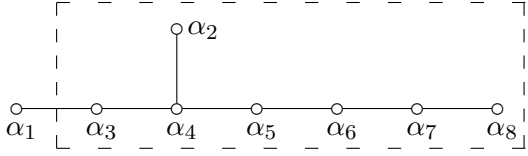
\begin{figure}[t]
\begin{picture}(400,50)
\put(100,20){\begin{picture}(0,0)
	
	\multiput(0,0)(30,0){7}{\circle{4}}
	\multiput(2,0)(30,0){6}{\line(1,0){26}}
	\put(-4,-10){$\alpha_1$}
	\put(26,-10){$\alpha_3$}
	\put(56,-10){$\alpha_4$}
	\put(86,-10){$\alpha_5$}
	\put(116,-10){$\alpha_6$}
	\put(146,-10){$\alpha_7$}
	\put(176,-10){$\alpha_8$}
	\put(60,30){\circle{4}}
	\put(60,2){\line(0,1){26}}
	\put(64,28){$\alpha_2$}

	\multiput(15,-15)(15.5,0){12}{\line(1,0){5}}
	\multiput(15,40)(15.5,0){12}{\line(1,0){5}}
	\multiput(15,-15)(0,10){6}{\line(0,1){5}}
	\multiput(190,-15)(0,10){6}{\line(0,1){5}}
	
	\end{picture}}
\end{picture}
\caption{
The Dynkin diagram of $E_8$ and its subdiagram corresponding to SO(14) $\subset E_8$.
}
\label{fig:DynkinE8}
\end{figure}
Let us now turn to the algebraic description of $E_2$ and its SO(2)$\times$U(1)$_I$ subgroup.
For $n\ge 3$, the Lie algebra $E_n$ corresponds to the Dynkin diagram of $E_8$ with simple roots $\alpha_{n+1},\dots,\alpha_8$ deleted, referring to the root labeling as in \figref{fig:DynkinE8}.\footnote{For $n=3,4$, the simple roots are relabeled as

\begin{picture}(400,40)
\put(60,17.5){\large$E_3:$}
\put(100,20){\begin{picture}(0,0)
	\multiput(0,0)(30,0){3}{\circle{4}}
	\multiput(2,0)(30,0){1}{\line(1,0){26}}
	\put(-4,-10){$\alpha_1$}
	\put(26,-10){$\alpha_2$}
	
	\put(56,-10){$\alpha_3$}
		
	\multiput(15,-15)(10,0){6}{\line(1,0){5}}
	\multiput(15,10)(10,0){6}{\line(1,0){5}}
	\multiput(15,-15)(0,10){3}{\line(0,1){5}}
	\multiput(70,-15)(0,10){3}{\line(0,1){5}}
	
	\end{picture}}
	
\put(220,17.5){\large$E_4:$}
\put(260,20){\begin{picture}(0,0)
	\multiput(0,0)(30,0){4}{\circle{4}}
	\multiput(2,0)(30,0){3}{\line(1,0){26}}
	\put(-4,-10){$\alpha_1$}
	\put(26,-10){$\alpha_2$}
	\put(56,-10){$\alpha_3$}
	\put(86,-10){$\alpha_4$}

	\multiput(15,-15)(10,0){9}{\line(1,0){5}}
	\multiput(15,10)(10,0){9}{\line(1,0){5}}
	\multiput(15,-15)(0,10){3}{\line(0,1){5}}
	\multiput(100,-15)(0,10){3}{\line(0,1){5}}
	
	\end{picture}}

\end{picture}}
This definition, however, is inadequate for $n=2$, as $E_2\cong$ SU(2)$\times$U(1) (and not SU(2)$\times$SU(2), as the extension of the above definition to $n=2$ might suggest). 
Before we proceed to the definition of $E_2$, let us list for reference the simple weights of $E_8$,
\bear
\Lambda_1 &=& 4\alpha_1+5\alpha_2+7\alpha_3+10\alpha_4+8\alpha_5+6\alpha_6+4\alpha_7+2\alpha_8,\nn\\
\Lambda_2 &=& 5\alpha_1+8\alpha_2+10\alpha_3+15\alpha_4+12\alpha_5+9\alpha_6+6\alpha_7+3\alpha_8,\nn\\
\Lambda_3 &=& 7\alpha_1+10\alpha_2+14\alpha_3+20\alpha_4+16\alpha_5+12\alpha_6+8\alpha_7+4\alpha_8,\nn\\
\Lambda_4 &=& 10\alpha_1+15\alpha_2+20\alpha_3+30\alpha_4+24\alpha_5+18\alpha_6+12\alpha_7+6\alpha_8,\nn\\
\Lambda_5 &=& 8\alpha_1+12\alpha_2+16\alpha_3+24\alpha_4+20\alpha_5+15\alpha_6+10\alpha_7+5\alpha_8,\nn\\
\Lambda_6 &=& 6\alpha_1+9\alpha_2+12\alpha_3+18\alpha_4+15\alpha_5+12\alpha_6+8\alpha_7+4\alpha_8,\nn\\
\Lambda_7 &=& 4\alpha_1+6\alpha_2+8\alpha_3+12\alpha_4+10\alpha_5+8\alpha_6+6\alpha_7+3\alpha_8,\nn\\
\Lambda_8 &=& 2\alpha_1+3\alpha_2+4\alpha_3+6\alpha_4+5\alpha_5+4\alpha_6+3\alpha_7+2\alpha_8.\nn
\eear
They satisfy $\innerP{\Lambda_i}{\alpha_j}=\delta_{ij}$, where $\innerP{\cdot}{\cdot}$ is the bilinear form on the root lattice.

The correct definition of our Lie algebra $E_n$, valid for $2\le n\le 8$, is as follows. First, consider the sublattice $\latQ_{8-n}$ of the $E_8$ root lattice that is generated by the simple roots $\alpha_{n+1},\dots,\alpha_8$. The root spaces of those roots of $E_8$ that are in $\latQ_{8-n}$ generate an su$(9-n)$ subalgebra.
Indeed, the roots $\alpha_{n+1},\dots,\alpha_8$ form a subdiagram of Dynkin type $A_{8-n}$.
The exponent of this subalgebra is a subgroup SU($9-n$) $\subset E_8$, and $E_n$ is defined as the commutant of this subgroup. Note that with this definition, the simple roots of $E_n$ are {\bf not} $\alpha_1,\dots,\alpha_n$. For example, for $E_7$, the root spaces of $\pm\alpha_8$ generate an su(2) subalgebra that does not commute with the root space of $\alpha_7$, so $\alpha_7$ cannot be a simple root of $E_7$, as defined. Instead, we define
$$
\alpha_7'\equiv\Lambda_7-\Lambda_6=
-2\alpha_1
-3\alpha_2
-4\alpha_3
-6\alpha_4
-5\alpha_5
-4\alpha_6
-2\alpha_7 
-\alpha_8\,,
$$
and the simple roots of $E_7$ can then be taken as $\alpha_1,\dots,\alpha_6,\alpha_7'$.
It is easy to verify that their inner products correspond to the Dynkin diagram of $E_7$, and they all have zero inner product with $\alpha_8$.
Similarly, for $n=5,6$, we take the simple roots of $E_n\subset E_8$ to be $\alpha_1,\dots,\alpha_{n-1},\Lambda_n-\Lambda_{n-1}$. For $n=4$, we take the simple roots of $E_4\subset E_8$ to be $\alpha_1, \alpha_2, \alpha_3, \Lambda_4-\Lambda_3-\Lambda_2$. For $n=3$ we take the simple roots of $E_3\subset E_8$ to be $\alpha_1, \Lambda_3-\Lambda_1-\Lambda_2, \Lambda_3-\Lambda_2$.

The case $E_2\subset E_8$ requires a more careful treatment. $E_2\simeq$ SU(2)$\times$U(1)${}_I$ is defined as the subgroup that commutes with the SU(7) $\subset E_8$ generated by the root spaces of $\alpha_3,\dots,\alpha_8$.
Define the root
$$
\beta\equiv\Lambda_2-\Lambda_1=
\alpha_1
+3\alpha_2
+3\alpha_3
+5\alpha_4
+4\alpha_5
+3\alpha_6
+2\alpha_7
+\alpha_8
\,.
$$
Then $\pm\beta$ are the only roots of $E_8$ that are orthogonal to $\alpha_3,\dots,\alpha_8$. The root spaces of $\beta$ and $-\beta$ generate an su(2) subalgebra whose exponent we identify with the SU(2) factor of $E_2$.
The intersection of this su(2) with the Cartan subalgebra of $E_8$ is spanned by $\beta^\star$, which is the element of the Cartan subalgebra that assigns to a state with weight $\lambda$ the charge $Q'(\lambda)\equiv\innerP{\beta}{\lambda}$.
Then, the generator of the U(1) factor of $E_2\cong$ SU(2)$\times$U(1) is $\gamma^\star$, with
$$
\gamma\equiv 3\Lambda_1-\Lambda_2
=7\alpha_1+7\alpha_2+11\alpha_3+15\alpha_4+12\alpha_5+9\alpha_6+6\alpha_7+3\alpha_8
\,,
$$
which is the unique (up to multiplication) element of the root lattice that is orthogonal to $\alpha_3,\dots,\alpha_8$ and $\beta$.\footnote{$\gamma$ is not a root because $\innerP{\gamma}{\gamma}=14$.}
Under the subgroup $E_2\times$SU(7) $\subset E_8$ the representation $\rep{248}$ decomposes as
\bear
\lefteqn{
\rep{248}=
(\rep{1},\rep{1})_0
+(\rep{3},\rep{1})_0
+(\rep{1},\rep{48})_0
+(\rep{1},\rep{7})_{4}
+(\rep{1},\overline{\rep{7}})_{-4}
}
\nn\\
&&
+(\rep{2},\rep{7})_{-3}
+(\rep{2},\overline{\rep{7}})_{3}
+(\rep{1},\rep{35})_{-2}
+(\rep{1},\overline{\rep{35}})_{2}
+(\rep{2},\rep{21})_1
+(\rep{2},\overline{\rep{21}})_{-1}\,.
\nn
\eear
We now define ``fugacities'' $y$ and $z$, so that the contribution of a hypothetical state with $E_8$ weight $\lambda$ (assumed to be orthogonal to $\alpha_3,\dots,\alpha_8$) to the $E_2\cong$ SU(2)$\times$U(1) index will be
$
y^{\innerP{\beta}{\lambda}}
z^{\tfrac{1}{7}\innerP{\gamma}{\lambda}}
\,.
$

Now, consider the U(1)${}_I\times$SO(2) $\subset E_2$ subgroup. For a state associated to a weight $\lambda$ of $E_8$, we can associate U(1)${}'\times$U(1) $\subset$ SU(2)$\times$U(1) $\cong E_2$ charges
$$
Q'(\lambda)=\innerP{\beta}{\lambda}
\,,\qquad
\widetilde{Q}(\lambda)=\tfrac{1}{7}\innerP{\gamma}{\lambda}.
$$
Then their SO(2) and U(1)${}_I$ charges are given by
$$
Q_1(\lambda)=\tfrac{1}{2}Q'(\lambda)+\tfrac{7}{2}\widetilde{Q}(\lambda)=
\innerP{\tfrac{1}{2}\beta+\tfrac{1}{2}\gamma}{\lambda}=\innerP{\Lambda_1}{\lambda}
\,,
$$
and
$$
Q_I(\lambda)=\tfrac{1}{2}Q'(\lambda)-\tfrac{1}{2}\widetilde{Q}(\lambda)
\equiv
\tfrac{1}{7}\innerP{\delta}{\lambda},
$$
where we defined
$$
\delta\equiv\tfrac{7}{2}\beta-\tfrac{1}{2}\gamma
=7\alpha_2+5\alpha_3+10\alpha_4+8\alpha_5+6\alpha_6+4\alpha_7+2\alpha_8,
$$
which is the (unique up to multiplication) weight that is orthogonal to $\Lambda_1$ and $\alpha_3,\dots,\alpha_8$.
As was discovered in \cite{Kim:2012gu}, the superconformal index for local operators is
$$
{\cal I}_{\text{\tiny SCI}} = 
1+(2 + \frac{1}{q y_1} + q y_1) t^2 + O(t^3)
= 1 + \left\lbrack
1 + \left(\frac{1}{y^2} + 1 + y^2\right)
\right\rbrack t^2 + O(t^3),
$$
and as we have seen in \secref{subsec:IndexRays}, the ray operator index is
$$
{\cal I}_{\text{\tiny ray}} = q^{-2/7}\left(\frac{1}{y_1^2}+\frac{q}{y_1}+y_1^2\right)t + O(t^2)
= 
\left\lbrack
z^{-3/7}\left(y+\frac{1}{y}\right) + z^{4/7}
\right\rbrack t + O(t^2),
$$
which both nicely fit into $E_2\cong$ SU(2)$\times$U(1) representations.
\end{appendix}


\bibliography{refs} 
\bibliographystyle{JHEP}

\end{document}